\documentclass[aps,prl,twocolumn,amsmath,amssymb,tightenlines,epsfig,floatfix,superscriptaddress]{revtex4-1}
\usepackage{graphicx}
\usepackage{dcolumn}	% Align table columns on decimal point
\usepackage{bm}			% bold math
\usepackage{amsfonts}
\usepackage{xspace}
\usepackage{color}
\usepackage{epstopdf}
\usepackage{multirow}
\usepackage{siunitx}

\begin{document}

\newcommand{\sss}[1]{{\color{blue}#1}}

\title{Thermally-robust spin correlations between two $^{85}$Rb atoms in an optical microtrap}

\author{Pimonpan Sompet}
\affiliation{The Dodd-Walls Centre for Photonic and Quantum Technologies, Department of Physics, University of Otago, Dunedin, New Zealand}
\author{Stuart S. Szigeti}
\affiliation{The Dodd-Walls Centre for Photonic and Quantum Technologies, Department of Physics, University of Otago, Dunedin, New Zealand}
\affiliation{Department of Quantum Science, Research School of Physics and Engineering, The Australian National University, Canberra ACT 2601, Australia}
\author{Eyal Schwartz}
\affiliation{The Dodd-Walls Centre for Photonic and Quantum Technologies, Department of Physics, University of Otago, Dunedin, New Zealand}
\author{Ashton S. Bradley}
\affiliation{The Dodd-Walls Centre for Photonic and Quantum Technologies, Department of Physics, University of Otago, Dunedin, New Zealand}
\author{Mikkel F. Andersen}
\affiliation{The Dodd-Walls Centre for Photonic and Quantum Technologies, Department of Physics, University of Otago, Dunedin, New Zealand}

\begin{abstract}

The complex collisional properties of atoms fundamentally limit investigations into a range of processes in many-atom ensembles. In contrast, the bottom-up assembly of few- and many-body systems from individual atoms offers a controlled approach to isolating and studying such collisional processes. Here, we use optical tweezers to individually assemble pairs of trapped $^{85}$Rb atoms, and study the spin dynamics of the two-body system in a thermal state. The spin-2 atoms show strong pair correlation between magnetic sublevels on timescales exceeding one second, with measured relative number fluctuations $11.9\pm0.3$~dB below quantum shot noise, limited only by detection efficiency. Spin populations display relaxation dynamics consistent with simulations and theoretical predictions for $^{85}$Rb spin interactions, and contrary to the coherent spin waves witnessed in finite-temperature many-body experiments and zero-temperature two-body experiments. Our experimental approach offers a versatile platform for studying two-body quantum dynamics and may provide a route to thermally-robust entanglement generation.

\end{abstract}

\maketitle

%\newpage
%%%%%%%%%%%%   Start: Introduction  %%%%%%%%%%%%%%
%\section*{Introductional Paragraph (abstract)}

\section*{Introduction}
When two atoms collide their interaction is complex, leading to a wide range of possible outcomes. The result of the collision strongly depends upon experimental parameters such as the internal atomic states, the collisional energy, and external electromagnetic fields~\cite{Horvath2017}. Modern atomic physics experiments exploit the richness of these atomic interactions to engineer systems for a remarkable variety of purposes, including quantum information processing~\cite{Isenhower2010} and quantum simulation~\cite{Bloch2012, Greiner2016}. A wealth of physical phenomena have been simulated with cold atoms, such as black holes~\cite{Lahav2010} and superconductivity~\cite{Greiner2003}. Of particular importance to atomic simulations of quantum magnetism is the local spin-changing interaction between atoms in their groundstate manifold~\cite{Stamper-Kurn2016, Williamson2017}.

In many-body experiments, spin-changing collisions lead to coherent spin waves in both quantum-degenerate and thermal atomic samples~\cite{Sengstock2004, Chapman2005, Lett2013, Ebling2014, Krauser:2014, He:2015, Hoang:2016}. These spin waves manifest as time-dependent populations of the atoms' magnetic sublevels. Spin-changing collisions have additionally been used to generate quantum-entangled samples of ten thousand atoms~\cite{You2017}. Such entanglement has enabled sub-shot-noise phase measurements with matter-wave interferometers~\cite{Klempt2011, Linnemann:2016} and has recently allowed fundamental studies of Einstein-Podolsky-Rosen (EPR) steering with atomic clouds~\cite{Fadel:2018, Kunkel2018, Lange:2018}.

Unfortunately, detailed investigations of spin-changing collisions in many-atom experiments is challenging, due to undesirable processes including three-body loss~\cite{Grimm2006,Salomon2017}. The superfluid to Mott insulator transition provides one means of separating atomic pairs for `clean' studies of spin-changing collisions~\cite{Bloch2005, Widera:2006}. However, this is limited to atomic species with collisional properties suitable for Bose condensing and subsequent manipulation. Consequently, experimental tests of the predicted $^{85}$Rb spin-dependent interaction strengths~\cite{Greene2001} have remained elusive, and in general atomic species with negative background scattering lengths suffer unique experimental difficulties in the many-body regime~\cite{Stamper-Kurn2016}.

A more versatile, bottom-up approach~\cite{Endres2016, Barredo:2016} is to prepare and manipulate individual atomic pairs via optical tweezers, enabling studies of interactions between any combination of atoms that can be laser cooled. However, to date such studies have been restricted to inelastic interactions that cause atom loss~\cite{Andersen2010, Andersen2013, Xu2015, Liu2018}, and interactions where no overall population dynamics occur~\cite{Regal2015}.

Here, we study spin-changing collisions between individual pairs of $^{85}$Rb atoms prepared in an optical tweezer, and observe the collision-driven population dynamics of the magnetic sub-states in the groundstate manifold. We observe record-high suppression of relative number fluctuations and find that a bias magnetic field strongly affects the dynamics. The observed crossover from fast relaxation dynamics at low-bias field to slow, field-independent relaxation dynamics at higher fields is captured by simulations based upon a simplified atom-atom interaction. However, for high magnetic fields the very large system of coupled modes involved at the experimental temperature prohibits quantitative first-principles modelling of the observed slow relaxation of spin-state populations. Nonetheless, in this regime the experimental data is well-fitted using incoherent rate equations with a single-parameter fit, where the relative coupling rates between different spin states is deduced from the theoretically-predicted $^{85}$Rb spin-dependent interaction strengths~\cite{Greene2001}.

%%%%%%%%%%% Figure 1 %%%%%%%%%%%%
\begin{figure}
	\begin{center}
     \includegraphics[width=\linewidth]{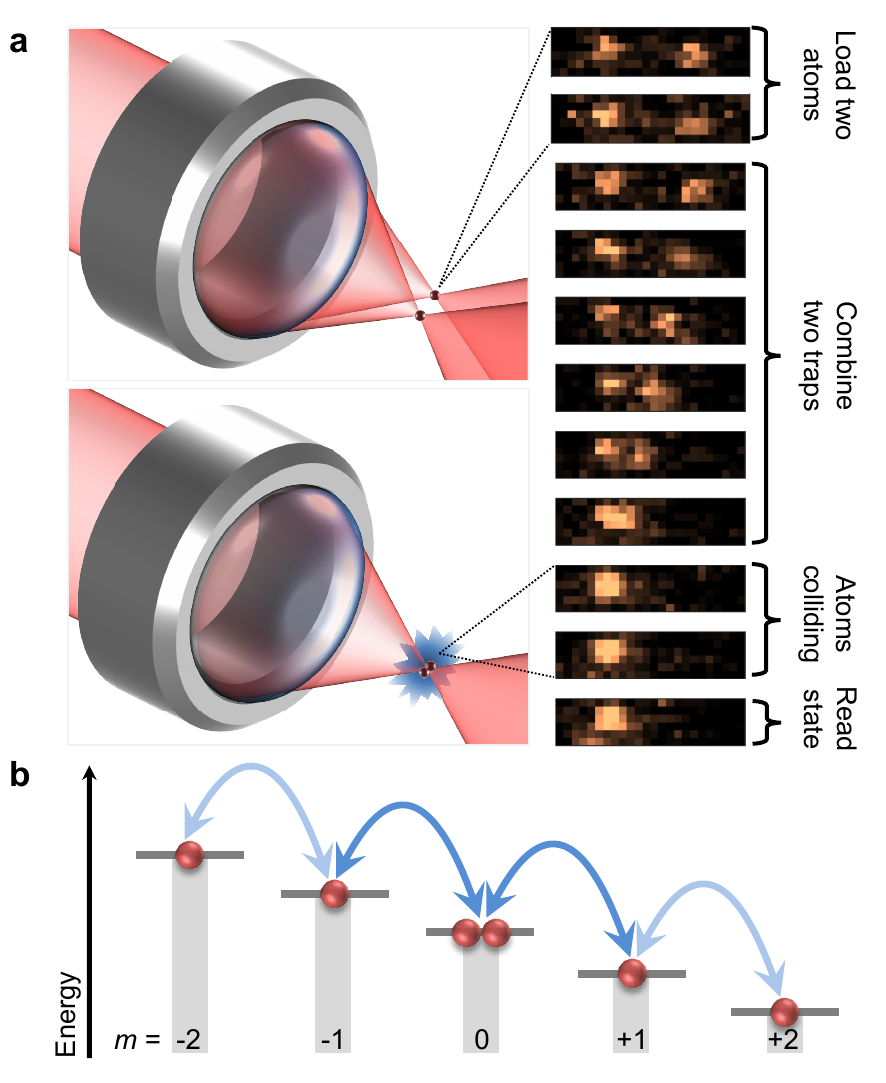}
	\end{center}
   	\caption{\textbf{Experimental schematics. a,} (Left) Two optical tweezers are formed using the high-numerical-aperture lens.
	By reducing the separation between the tweezers and then turning one of the tweezers off, both atoms are transferred into the same optical tweezer, allowing atomic collisions to occur.
	(Right) Superimposed fluorescent images of the same two atoms showing their relative positions for different experimental stages.
	After combining the two traps, the individual atomic positions can no longer be resolved.
	\textbf{b,} Spin-changing collisions: Two atoms initially in $\left|0,0\right\rangle$ can only couple to $\hat{S}\left|1,-1\right\rangle$ (dark arrows) and then to $\hat{S}\left|2,-2\right\rangle$ (light arrows), where the symmetrization operator $\hat{S}$ is defined in the main text.
	}

	%\vspace{0cm}
	\label{fig:figure1}
	\end{figure}
%%%%%%%%%%%%%%%%%%%%%%%%%%%%%%

\section*{Results}
\noindent \textbf{Experimental Sequence. }
Our experiments employ two $^{85}\mathrm{Rb}$ atoms, initially loaded into two separated optical tweezers~\cite{Andersen2013_2,Andersen2010,Regal2015a}, and prepared in the $f=2$, $m=0$ groundstate (see Fig.~\ref{fig:figure1}a). 
The two optical tweezers are then merged, leaving the pair in a single tweezer. The magnetic bias field is set to the desired value and the two atoms are held within the single tweezer for a specified duration, which we hereafter refer to as the collision time. After a given collision time, the atomic $m$-states (denoted $\left|m\right\rangle$) are measured by ejecting atoms in a particular $\left|m\right\rangle$ and measuring the remaining atom number (see Methods for details and experimental parameters). \\
%%%%%%%%%%%%   End: Experimental sequence  %%%%%%%%%%%%%%

\noindent \textbf{Model.}
Once in the same optical tweezer, the two atoms interact via interaction Hamiltonian $\hat{H}_\textrm{s}$, which depends on the pair's relative position and spin state. Approximating the optical tweezer as an $m$-independent harmonic potential separates the centre-of-mass and relative motions of the two atoms, decoupling the internal spin and centre-of-mass dynamics, and permitting a simplified description via Hamiltonian~\cite{Tiesinga:2000, Blume:2002, Bloch2005, Widera:2006}
\begin{equation}
 \hat{H} = \frac{\hat{\textbf{p}}^2}{2\mu} + \sum_{j=x,y,z}\frac{1}{2}\mu \omega_{j}^2 \hat{r}_{j}^2 + \sum_{i = 1,2}\hat{H}_{\textrm{Z},i} + \hat{H}_\textrm{s}, \label{Hamiltonian}
\end{equation}
where $\hat{\textbf{r}} = (\hat{r}_x,\hat{r}_y,\hat{r}_z)$ and $\hat{\textbf{p}}$ are relative position and momentum operators, respectively, $\mu$ the reduced mass, $\omega_j$ the atomic oscillation frequency in
the $j^\text{th}$ dimension, and $\hat{H}_{\textrm{Z},i}$ the Zeeman shift for the $i^{\text{th}}$ atom. Our experiments use thermal atoms with $k_\textrm{B}T$ much larger than $\hbar \omega_j$, Zeeman energies, and atomic interaction energies.

Under suitable approximations, $\hat{H}_\textrm{s}$ conserves total magnetization~\cite{Stamper-Kurn2013, Bloch2005, Widera:2006} and two atoms initially prepared in $m_1=m_2=0$ are restricted to bosonic symmetrized states with $m_1=-m_2$: $\left|0,0\right\rangle={\left|0\right\rangle}_1\otimes{\left|0\right\rangle}_2$, $\hat{S}\left|1,-1\right\rangle=\frac{1}{\sqrt{2}}\left({\left|1\right\rangle}_1\otimes{\left|-1\right\rangle}_2+{\left|-1\right\rangle}_1\otimes{\left|1\right\rangle}_2\right)$, and $\hat{S}\left|2,-2\right\rangle= \frac{1}{\sqrt{2}}\left({\left|2\right\rangle}_1\otimes{\left|-2\right\rangle}_2+{\left|-2\right\rangle}_1\otimes{\left|2\right\rangle}_2\right)$ (see Fig.~\ref{fig:figure1}b). Here $\hat{S}$ denotes the symmetrization operator, $\left|m_1, m_2 \right\rangle$ the unsymmetrized two-particle spin states, and subscripts 1 and 2 denote the two atoms. \\

\begin{figure}
	\begin{center}
     \includegraphics[width=\linewidth]{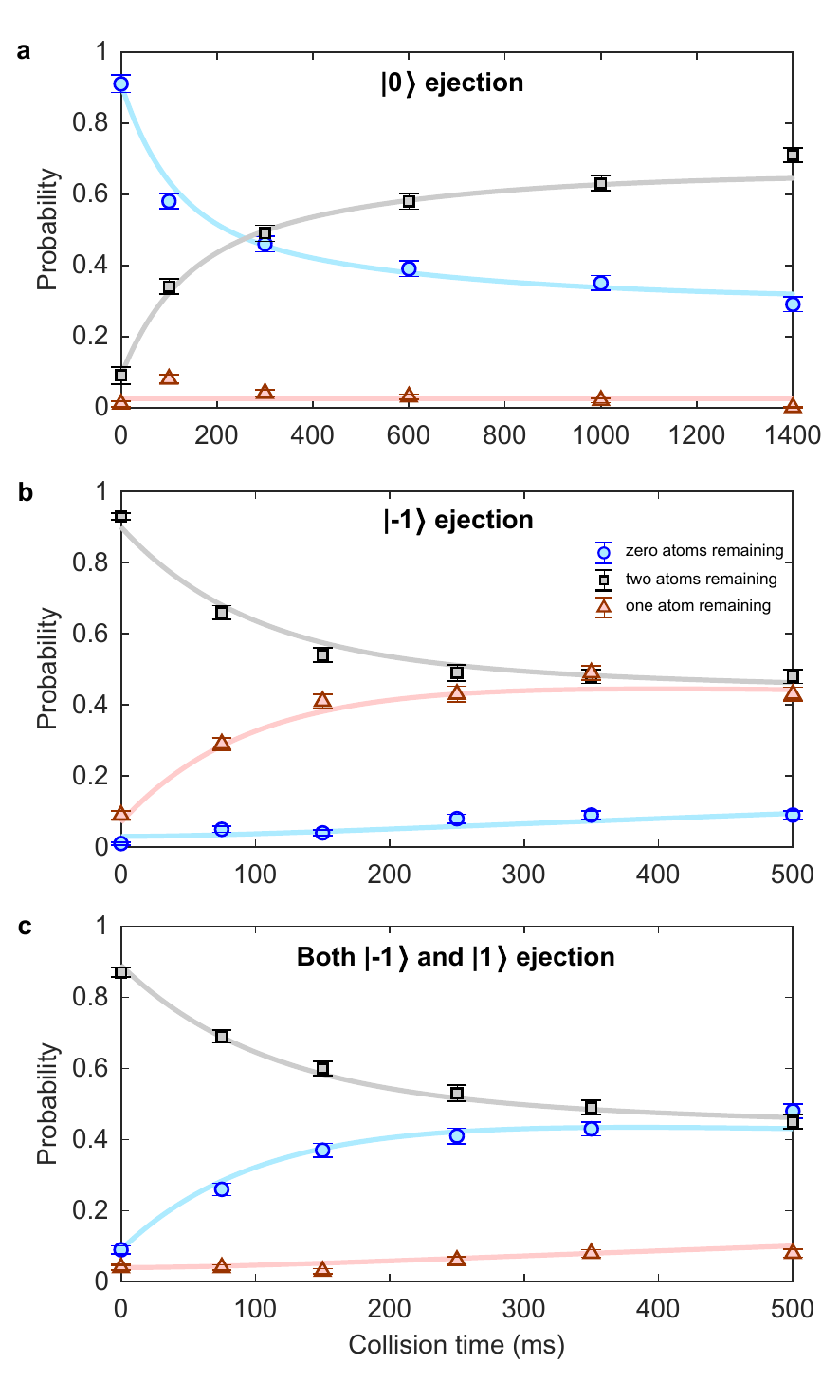}
	\end{center}
	\vspace{-1cm}
   	\caption{
	{\bf \boldsymbol{$m$}-state correlation results.}
	 Probability that zero, one, or two atoms remain in the optical tweezer after a given collision time.  \textbf{a,} When atoms in $\left|0\right\rangle$ are expelled (immediately after a given collision time), the probability that both atoms were in $|0\rangle$ (and therefore ejected) decreases, while the probability that both atoms remain correspondingly increases.
	\textbf{b,} Expelling atoms solely from $\left|-1\right\rangle$ gives only single-atom loss events.
	\textbf{c,} Expelling atoms from both $\left|-1\right\rangle$ and $\left|1\right\rangle$ gives only pair loss, in strong contrast to the result in \textbf{b}.
	In all cases and throughout the collision time, the bias magnetic field was 8.5 Gauss. Error bars in all panels denote the standard error of the mean. The solid curves are fits to the data included to guide the eye. Similar data that demonstrates correlations between $|-2\rangle$ and $|2\rangle$ is shown in Supplementary Fig.~1. Source data are provided as a Source Data file.}
%	}
	%\vspace{0cm}
	\label{fig:figure2}
\end{figure}

\noindent \textbf{Spin correlations.}
By measuring magnetic sublevels of the atomic pair for different collision times, we confirm that the spin dynamics is governed by the simple model of spin-changing collisions depicted in Fig.~\ref{fig:figure1}b, which yields strong correlations between the $m$-states in a given pair. This requires the three measurement series summarized in Fig.~\ref{fig:figure2}. A particular $\left|m\right\rangle$ is detected by ejecting atoms in this state. In Fig.~\ref{fig:figure2}a we expel atoms in $\left|0\right\rangle$ after a given collision time. The probability that both atoms are in $\left|0\right\rangle$ (i.e. no remaining atoms) decays with increasing collision time, while the probability that both atoms remain grows correspondingly. The probability of observing one remaining atom is always negligible, implying that collisions cause both atoms to leave $\left|0\right\rangle$ simultaneously. 
In Fig.~\ref{fig:figure2}b we start with both atoms in $\left|0\right\rangle$ but eject atoms in $|-1\rangle$. The probability that one atom is in $\left|-1\right\rangle$ grows with collision time, but both are never $\left|-1\right\rangle$, since the probability that both atoms are ejected is effectively zero. In Fig.~\ref{fig:figure2}c we eject atoms in both $\left|-1\right\rangle$ and $\left|1\right\rangle$. This ejects both atoms, or none. Combining this with Fig.~\ref{fig:figure2}b, we conclude that when one atom is in $\left|-1\right\rangle$, the other is in $\left|1\right\rangle$. The populations of $\left|-1\right\rangle$ and $\left|1\right\rangle$ are therefore almost perfectly correlated. Similar data for $\left|\pm2\right\rangle$ shows these populations are also correlated (see Supplementary Note~1). The lasting pair correlation on timescales exceeding one second is facilitated by having individual atomic pairs. In contrast, in many-body experiments with spin-2 atoms, subsequent spin-changing collisions would likely deteriorate such strong pair correlations.

We quantify the pair correlation with the relative number squeezing, $\zeta^2$ (see Methods). Without correcting for finite detection efficiency, it is $11.9\pm0.3$ dB below quantum shot noise (QSN) for the $\left| \pm 1 \right\rangle$ populations. Since our atomic-pair ensemble is thermal, this large pair correlation is thermally robust. $\zeta^2$ is limited solely by our detection efficiency (see Methods); improved detection efficiency could reduce $\zeta^2$ by a further order of magnitude. For many-body systems, the highest reported relative number squeezing via spin-changing collisions is 11.4 dB below QSN (12.4 dB after correcting for detection inefficiency)~\cite{Klempt2014}. \\

%%%%%%%%%%% Figure 2 %%%%%%%%%%%%
%%%%%%%%%%%%%%%%%%%%%%%%%%%%%%

\begin{figure}
	\begin{center}
     \includegraphics[width=1\linewidth]{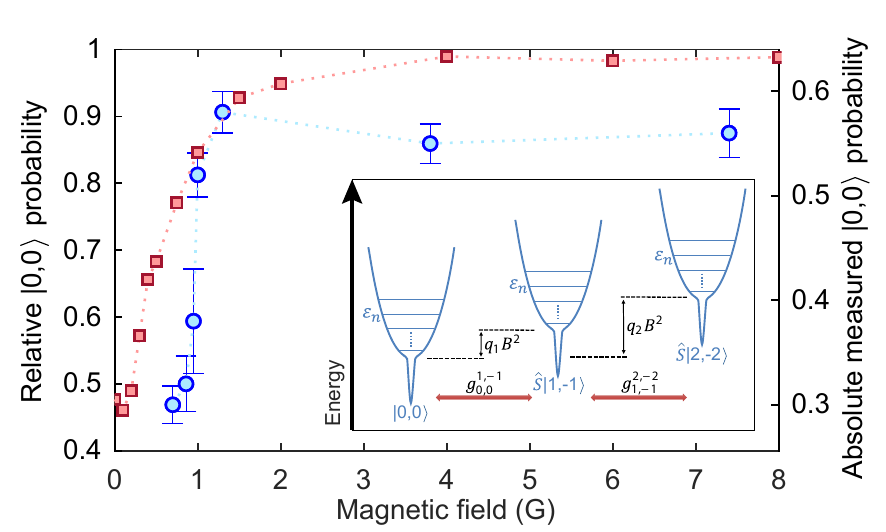}
	\end{center}
   	\caption{\textbf{Effect of bias magnetic field.} The left axis represents the $\left|0,0\right\rangle$ population at 40 ms of collision time relative to the $\left|0,0\right\rangle$ population at $t=0$. Both the experimental (blue circles) and simulation (red squares) results are plotted as a function of the magnetic field. Error bars in the experimental data denote the standard error of the mean. Although the $\left|0,0\right\rangle$ population of the simulation at $t=0$ is set to 1, in the experiment the population dynamics during the magnetic-field ramp leave a $\left|0,0\right\rangle$ population of about 0.64 at $t=0$. 
	The right axis is the actual scale of the experimentally-measured $\left|0,0\right\rangle$ population. 
	The inset schematically shows the energy-level picture of the system. Atomic pairs in a given spin mode have accessible energies $\varepsilon_\textbf{n} = \hbar \omega_x (n_x+\tfrac{1}{2})+\hbar \omega_y (n_y+\tfrac{1}{2})+\hbar \omega_z (n_z+\tfrac{1}{2})$, constrained by $(-1)^{n_x+n_y+n_z} = 1$. A magnetic field of strength $B$ shifts the energy levels of modes $\hat{S}|1,-1\rangle$ and $\hat{S}|2,-2\rangle$ by $q_1 B^2$ and $q_2 B^2$, respectively, due to the quadratic Zeeman effect. Spin-changing collisions couple these energy levels, with coupling strengths $g_{m_1,m_2}^{m_1',m_2'}$. See Methods for further details. Source data are provided as a Source Data file.
	}
	%\vspace{0cm}
	\label{fig:figure3}
\end{figure}

\noindent \textbf{Magnetic field dependence.}
The bias magnetic field affects the spin dynamics through $\sum_{i}\hat{H}_{\textrm{Z},i}$. Since our model conserves total magnetization, the first-order Zeeman contributions cancel for the accessible two-body states, so $\sum_{i}\hat{H}_{\textrm{Z},i}$ only contributes via second-order terms. We investigate how $\sum_{i}\hat{H}_{\textrm{Z},i}$ affects the spin dynamics by measuring the $\left|0,0\right\rangle$ population after a 40$\,$ms collision time for different bias fields (Fig.~\ref{fig:figure3}). At low biases, the dynamics are highly magnetic-field dependent, whereas for higher biases the dynamics are effectively magnetic-field independent.
Here typical thermal energies are much larger than second-order Zeeman energies for all biases investigated. The atom pairs therefore have sufficient thermal energy to overcome the Zeeman shift when undergoing spin-changing collisions, so, in contrast to ultracold samples, the dynamics should not necessarily quench at high biases. 

To understand the spin evolution, we simulated the dynamics governed by Eq.~(\ref{Hamiltonian}) with a simplified interaction $\hat{H}_\textrm{s} = V\left(\hat{\textbf{r}}\right)\times\sum_{m_1, m_2, m'_1, m'_2}g_{m_1,m_2}^{m'_1,m'_2} \left|m'_1, m'_2 \right\rangle \left\langle m_1,m_2\right|$,
where $g_{m_1,m_2}^{m'_1,m'_2}$ are determined from predicted spin-dependent $s$-wave scattering lengths~\cite{Greene2001} and $V\left(\hat{\textbf{r}}\right)$ is a Gaussian with width chosen to reproduce the total free-space $s$-wave collision cross section (see Methods). A Gaussian pseudopotential moderates problems that afflict zero-length interaction potentials in tight traps~\cite{Tiesinga:2000, Blume:2002}, while still avoiding the complexity of a more complete $\hat{H}_\textrm{s}$.

The simulation was conducted by averaging over a thermal ensemble of initial states evolved using Eq.~(\ref{Hamiltonian}). The initial states were relative-motion eigenstates of $\hat{\textbf{p}}^2 / (2\mu) + \sum_{j}\tfrac{1}{2}\mu \omega_{j}^2 \hat{r}_{j}^2$ with two-particle spin state $\left|0,0\right\rangle$. Due to the prohibitively-large Hilbert space required at the experimental temperature, simulations were restricted to a lower temperature of $\SI{8.8}{\micro\kelvin}$. All simulations at this temperature were performed on a finite basis of 16,996 relative-motional modes.

The simulation qualitatively captures the spin dynamics (Fig.~\ref{fig:figure3}). We observe a crossover from fast dynamics at low magnetic-field strengths to slow dynamics at high fields. $\hat{H}$ couples the three allowed spin modes, $\left|0,0\right\rangle$, $\hat{S}\left|1,-1\right\rangle$, and $\hat{S}\left|2,-2\right\rangle$ (inset, Fig.~\ref{fig:figure3}). When the pair is in a particular spin mode, it behaves as an effective single particle within a harmonic trap with the interaction potential placed at the trap centre. At low magnetic fields, $\sum_{i}\hat{H}_{\textrm{Z},i}$ is negligible, so any relative-motion eigenstate with a particular spin mode (e.g. $\left|0,0\right\rangle$) is approximately degenerate to relative-motion eigenstates in other spin modes (e.g. $\hat{S}\left|1,-1\right\rangle$, and/or $\hat{S}\left|2,-2\right\rangle$); the degeneracy is only lifted by the atom-atom interaction's spin-state dependence.
The resulting resonant coupling efficiently transfers population between spin modes at low magnetic fields. In contrast, at high fields this degeneracy is lifted, the majority of initially-occupied states have no near-resonant coupling to other spin modes, leaving only off-resonant coupling, and the dynamics largely cease. \\

%%%%%%%%%%% Figure 3 %%%%%%%%%%%%
%%%%%%%%%%%%%%%%%%%%%%%%%%%%%%

\begin{figure}
	\begin{center}
     \includegraphics[width=1\linewidth]{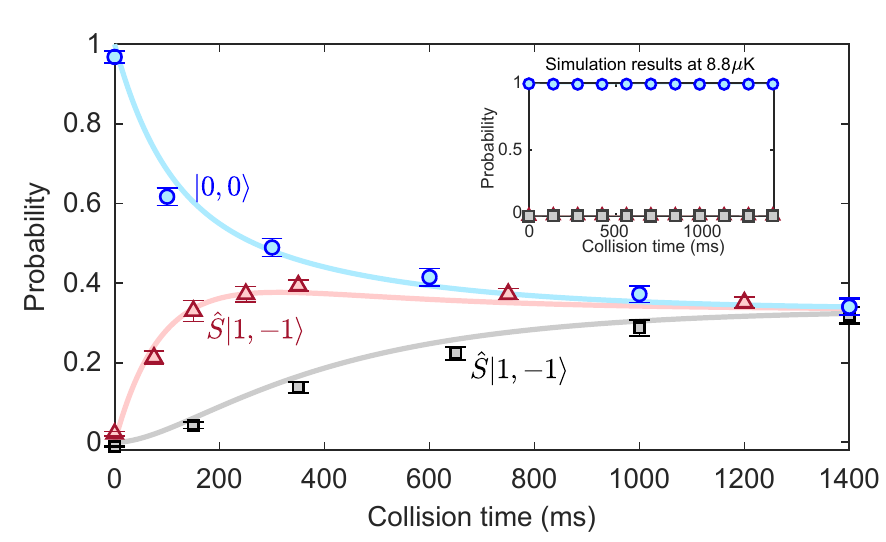}
	\end{center}
   	\caption{\textbf{Spin population dynamics at high bias field.} The populations of the two-atom states are plotted as a function of collision time with error bars denoting the standard error of the mean.
	The solid curves are a fit of the measured data with spin-changing rate equations, while the ratio of the rates between $\left|0,0\right\rangle\rightleftharpoons\hat{S}\left|1,-1\right\rangle$ and $\hat{S}\left|1,-1\right\rangle\rightleftharpoons\hat{S}\left|2,-2\right\rangle$ is determined from the theoretically-predicted spin-dependent interaction strengths. The bias field was 8.5 Gauss for all collision times. The inset illustrates that the simplified theoretical model used for our simulations fails to capture the long-time relaxation dynamics in the high magnetic-field regime. Source data are provided as a Source Data file.
	}
	%\vspace{0cm}
	\label{fig:figure4}
	\end{figure}

\noindent \textbf{The high magnetic field regime.}
Figure~\ref{fig:figure3} shows a quantitative difference between simulation and experiment. In the high bias, magnetic-field-independent regime, the simulation gives $\left|0,0\right\rangle$ population at $t=40\,$ms close to the $t=0$ population, while in the experiment it is lower. Figure~\ref{fig:figure4} demonstrates the cause of this difference.
The experiment shows slow relaxation to equal populations of the three spin modes, while the simulation dynamics are quenched (no spin-changing collisions). Here equal population is not complete thermalization within states that conserve total magnetization; since atoms with different internal states can be considered distinguishable, the thermalized populations with $m=\pm1$ and $m=\pm2$ would be twice that of $|0,0\rangle$.

Generally, \emph{a priori} calculations of thermal decoherence in colliding atomic ensembles pose a challenge for theory, often necessitating phenomenological rate-equation approaches to account for dissipation~\cite{Oktel:2002, Williams:2002, Fuchs:2002, Bradley:2002}. In our system, several effects that are not included in the simulations might explain the dynamics in Fig.~\ref{fig:figure4}. Magnetic field noise might affect the dynamics or slight polarization pollution of the optical tweezer light could give a slightly $m$-dependent trap, the latter invalidating our separation of the pair's centre-of-mass and relative coordinates. The non-paraxial nature of the optical tweezers inevitably introduces a spatially-varying polarization that can be described as a fictitious magnetic field gradient~\cite{Thompson2013}. We suppress the effect of this by having the bias magnetic field perpendicular to the fictitious field. A more realistic atom-atom interaction $\hat{H}_\textrm{s}$ may also introduce new collisional timescales not captured by our simulations' simplified interaction. Finally, the five-fold temperature difference between our simulations' practical limit and the experimental temperature could play a role. However, this appears an unlikely explanation, as the simulation does not reveal long-time dynamics for any of the temperatures we investigated. Note that Refs~[~\cite{Bloch2005, Widera:2006}] also included fitted relaxation rates with timescales similar to what we observe in Fig.~\ref{fig:figure4}, and this was needed in order to match their experimental observations to theoretical predictions.

Figure~\ref{fig:figure4}'s data is well-modelled using rate equations (see Methods). Incoherent transition rates likely depend on the cross section for the process, which is proportional to the squared magnitude of the coupling matrix elements. These are determined from theoretically-predicted $^{85}$Rb spin-dependent interaction strengths~\cite{Greene2001}. Based on this, the ratio of the rates between $\left|0,0\right\rangle\rightleftharpoons\hat{S}\left|1,-1\right\rangle$ and $\hat{S}\left|1,-1\right\rangle\rightleftharpoons\hat{S}\left|2,-2\right\rangle$ is $2.34$, while the rate between $\left|0,0\right\rangle\rightleftharpoons\hat{S}\left|2,-2\right\rangle$ is negligible. Fitting using a single overall rate as the fitting parameter matches the data very well (Fig.~\ref{fig:figure4}), indicating that the ratios between the rates is determined by the ratios between the collisional cross sections. Figure~\ref{fig:figure3} therefore displays a crossover from a resonant coupling regime at low magnetic fields to a regime at high fields where the collision dynamics do not depend upon the energy difference between different spin states.
Although an incoherent rate equation model gives a good fit to the collisional dynamics in the high magnetic-field regime, it is incapable of providing an explanation of the magnetic-field dependence of the relaxation timescale in the low bias regime. The coupling matrix elements are independent of bias magnetic fields in the range we consider, and models that ignore quantization of the motional states do not capture the change in resonance condition that changing the bias field gives rise to. \\

%%%%%%%%%%% Figure 4 %%%%%%%%%%%%
%%%%%%%%%%%%%%%%%%%%%%%%%%%%%%

%%%%%%%%%%% Figure 5 %%%%%%%%%%%%
\begin{figure}
	\begin{center}
     \includegraphics[width=1\linewidth]{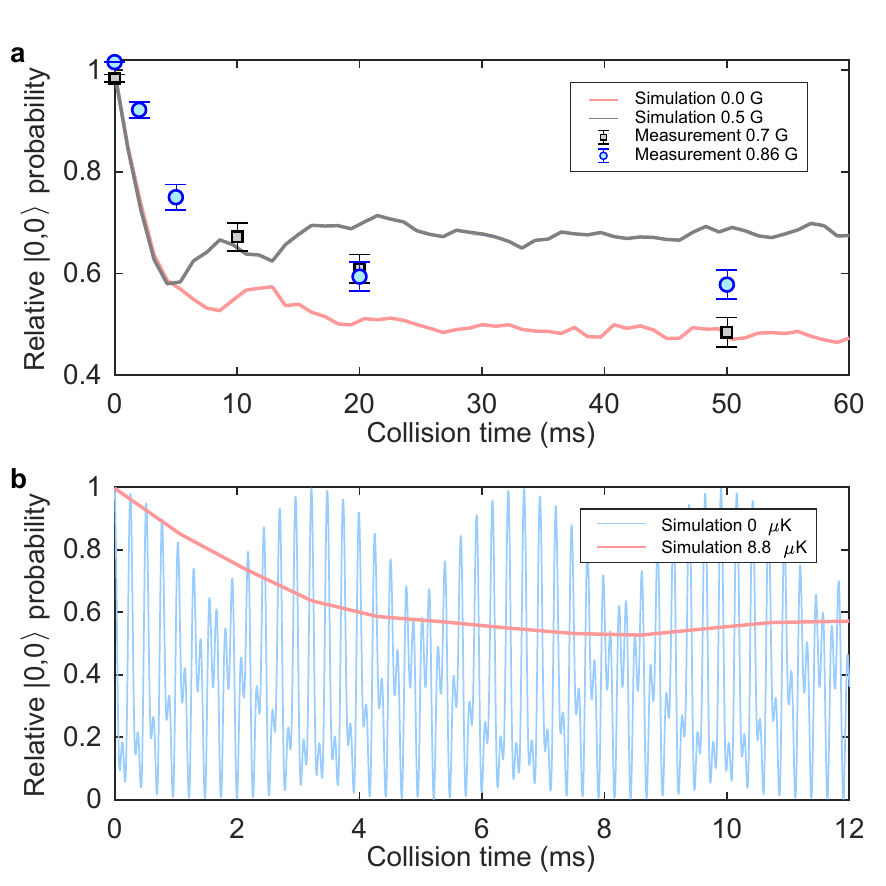}
	\end{center}
   	\caption{\textbf{Spin population dynamics at low bias field.} \textbf{a,} Comparison between simulated and measured relative populations of $\left|0,0 \right\rangle$ at low magnetic bias fields. Error bars in the experimental data denote the standard error of the mean. \textbf{b,} Simulation of the $\left|0,0 \right\rangle$ population as a function of collision time at two different temperatures and zero magnetic bias field. The initial relative motional state for the zero temperature simulation was the interacting groundstate of the relative motional Hamiltonian $\langle 0,0 | \hat{H} | 0,0\rangle$. The zero temperature simulation was performed on a truncated basis of 316 relative-motional modes. Source data are provided as a Source Data file.
	}
	%\vspace{0cm}
	\label{fig:figure5}
\end{figure}
%%%%%%%%%%%%%%%%%%%%%%%%%%%%%

\noindent \textbf{The low magnetic field regime.}
Figure~\ref{fig:figure5}a shows the measured and simulated $\left|0,0 \right\rangle$ populations as a function of collision time in the low bias-field regime. Both experimental data and simulation display spin relaxation dynamics. This is contrary to finite-temperature many-body experiments~\cite{Lett2013, He:2015} which exhibit high-contrast coherent oscillations between spin modes. The observed relaxation dynamics of the two-atom system can be understood from the form of the coupling matrix elements (that include the elements of $\mathbf{T}$, see Methods) that couple the different spin and relative motion states. Coupling between any two relative-motional eigenstates is strongly dependent upon the relative motional energies of these two states. They have a tendency to decrease as the relative motional energy increases, reflecting that the overall interaction decreases with increasing energy. Consequently, the timescale of the dynamics depends upon the initial relative motional state. Although each initial atom-pair state displays coherent oscillations, averaging over a thermal distribution of these initial states therefore washes out the oscillations, resulting in relaxation dynamics. This is illustrated in Fig.~\ref{fig:figure5}b, which shows a simulation of the $\left| 0,0 \right\rangle$ population for two different temperatures. At zero temperature, where only the relative motional groundstate is initially populated, we observe coherent oscillations similar to those in Ref.~[~\cite{Bloch2005}], while at $\SI{8.8}{\micro\kelvin}$ we see relaxation. Finally, since the coupling matrix elements decrease with increasing motional energy we also expect the lower temperature simulation to display faster dynamics than the experiment, consistent with Fig.~\ref{fig:figure5}a.  

\section*{Discussion}
Correlations alone is not evidence of entanglement. Nonetheless, from a theoretical perspective there should be entanglement in the spin sector despite the fact that we observe relaxation dynamics between the different spin states involved. Since $g_{0,0}^{1,-1}=g_{0,0}^{-1,1}$,  $\hat{H}_\textrm{s}$ only couples a pair initially in $\left|0,0\right\rangle$ to the symmetrized states $\hat{S}\left|1,-1\right\rangle$ and $\hat{S}\left|2,-2\right\rangle$, which are both entangled two-atom spin states. The interaction does not provide coupling to antisymmetrized spin states, for example $\hat{A} \left|1,-1\right\rangle \equiv \frac{1}{\sqrt{2}}\left({\left|1\right\rangle}_1\otimes{\left|-1\right\rangle}_2-{\left|-1\right\rangle}_1\otimes{\left|1\right\rangle}_2\right)$, since $\langle 1,-1 | \hat{A}^\dagger \hat{H}_\textrm{s} | 0, 0 \rangle=0$. Consequently, the collisional interaction alone does not provide a route for relaxation into unentangled two-atom spin states such as ${\left|-1\right\rangle}_1\otimes{\left|1\right\rangle}_2$ or ${\left|1\right\rangle}_1\otimes{\left|-1\right\rangle}_2$, since these are superpositions of $\hat{S}\left|1,-1\right\rangle$ and $\hat{A}\left|1,-1\right\rangle$. The relaxation dynamics that we observe in the theoretical calculations, consistent with the experiment at low magnetic bias fields, is therefore a relaxation into a mixture of $\left|0,0\right\rangle$, $\hat{S}\left|1,-1\right\rangle$, and $\hat{S}\left|2,-2\right\rangle$. Postselecting on any of the latter two entangled states therefore allows the preparation of a pure entangled state (see Supplementary Note~2).

Since ${\left|-1\right\rangle}_1\otimes{\left|1\right\rangle}_2$ is degenerate with ${\left|1\right\rangle}_1\otimes{\left|-1\right\rangle}_2$, unwanted effects such as magnetic field noise do not dephase $\hat{S}\left|1,-1\right\rangle$ into a mixture of unentangled states. Other effects such as spin-orbit coupling and polarization gradients from the non-paraxial nature of the optical tweezer, which are not presently included in our modelling, might also affect the quality of the entangled state. However, the strong correlation we observe between $m$-state populations justifies our neglect of spin-orbit coupling, and our choice of large trap detuning and alignment of the bias magnetic field perpendicular to the fictitious magnetic field mitigate the effects of polarization variations. We therefore expect that it should be possible to observe long-lived entanglement generated by the collisional interaction. Since states of the form $\frac{1}{\sqrt{2}}\left( \left|1,-1 \right\rangle + \left|-1,1 \right\rangle \right)$ have applications to metrology and quantum information processing~\cite{Pezze:2009, Nielsen:2010}, it is a future goal of ours to experimentally confirm the generation of the entangled state directly. For instance, exposing $\tfrac{1}{\sqrt{2}}(|1,-1\rangle + |-1,1\rangle)$ to a $\frac{\pi}{2}$-pulse (effected by driving stimulated Raman transitions between the $m = \pm1$ states) converts it to $-\tfrac{i}{\sqrt{2}}(|1,1\rangle + |-1,-1\rangle)$, which is identified by observing both atoms in the same $m$-state. If the entanglement was lost, we would observe both atoms in different $m$-states after the $\frac{\pi}{2}$-pulse with 50\% probability.

In the context of observing entanglement in our atom-pair system, we make two remarks on the experimental data from the high bias magnetic field regime where the relaxation mechanism is not yet captured by our simulations. First, we observe strong correlations between the two atoms' $m$-states in this regime, which is a requirement for entanglement. Secondly, Fig.~\ref{fig:figure4} does not show relaxation to equal populations of all five spin states that conserve total magnetization ($|0,0\rangle$, ${\left|-1\right\rangle}_1\otimes{\left|1\right\rangle}_2$, ${\left|1\right\rangle}_1\otimes{\left|-1\right\rangle}_2$, ${\left|-2\right\rangle}_1\otimes{\left|2\right\rangle}_2$, and ${\left|2\right\rangle}_1\otimes{\left|-2\right\rangle}_2$). Specifically, a $\chi$-squared test reveals that the observed relative populations at the final time point in Fig.~\ref{fig:figure4} significantly differ from $N_{m = 0} = \frac{1}{5}$ and $N_{m = \pm 1} = N_{m = \pm 2} = \frac{2}{5}$ ($\chi^2(\textrm{df}=3) = 71.1$, $p < 0.001$). In contrast, there is no statistically significant difference between these data and $N_{m = 0} = N_{m = \pm 1} = N_{m = \pm 2} = \frac{1}{3}$ ($\chi^2(\textrm{df}=3) = 1.9$, $p = 0.590$).
This could indicate that the antisymmetrized spin states remain unpopulated and entanglement is present in this regime. Although promising, these observations alone do not provide unequivocal evidence for entanglement in the spin sector.

\section*{Conclusions}
To summarize, the bottom-up assembling of pairs from individual atoms allows us to study the collisional properties of $^{85}$Rb, whose effective attractive interactions are unfavourable for ultracold-ensemble collision experiments. A single pair of $^{85}$Rb atoms in an optical tweezer displays spin dynamics that yield strong correlation between magnetic substates of the two atoms. Unlike both finite-temperature many-body experiments and zero-temperature two-body experiments, our finite-temperature two-body experiments show relaxation dynamics rather than coherent spin waves. The record-high pair correlation measured is only limited by detection inefficiency; improving upon this technical limitation might allow studies of unexplored effects, such as violations of total magnetization conservation due to spin-orbit coupling, or studies of quantum relaxation processes and quantum thermodynamics. Our experiments indicate that spin-changing collisions may offer a useful finite-temperature entanglement resource that is robust to thermal noise.

\section{Methods}
\noindent \textbf{Experimental procedure.}
We initially cool and trap a cloud of $^{85}$Rb atoms using magneto-optical trapping. We then load a small number of atoms from the cloud into two optical tweezers separated by $\sim \SI{4}{\micro \meter}$, each with a trap width of $\sim \SI{1.05}{\micro \meter}$  and depth of $h \times 58$ MHz. The two optical tweezers are formed by focusing two steerable linearly polarized laser beams ($\lambda=1064$ nm) with a high-numerical-aperture lens ($\text{NA}=0.55$). We use blue-detuned light-assisted collisions to reduce the occupancy of each trap to a single atom and confirm the presence of the two isolated atoms via fluorescence imaging~\cite{Andersen2010, Andersen2013_2, Regal2015a}. The probability that there are two atoms, one in each tweezer, after the loading procedure is $\sim$0.64, and we disregard the unsuccessful attempts.

After the loading process, the atoms are prepared in the desired $f=2$, $m=0$ groundstate in two steps. First, we optically pump atoms to the $f=3$, $m=0$ state by applying linearly-polarized optical pumping light with two frequencies corresponding to the  $^{85}$Rb $D_1$ $f=2$ to $f'=3$ and the $f=3$ to $f'=3$ transitions. During this, the bias magnetic field of 8.5 Gauss defines the quantization axis for the atoms in the groundstate. This gives an atomic population of 0.99 occupying the $f=3, m=0$ state. Last, we apply a $\pi$-pulse (1.57 $\mu$s) of co-propagating Raman beams ($\sim$36 GHz red detuned from the $D_2$ line) to coherently transfer the atoms from the $f=3, m=0$ state to the $f=2, m=0$ state.

Using a 20 ms frequency sweep of an acousto-optical modulator, we adiabatically bring the two tweezers closer until they are merged (the distance between the centres of the two laser beams is $\sim$900 nm). We then adiabatically ramp off one of the tweezers in $\sim$17 ms while the other is simultaneously ramped to the desired trap depth and the bias magnetic field is set to the chosen value.
The procedure leaves the atoms in the same optical tweezer where the collisional interactions generate the $\left|m\right\rangle$ population dynamics.

To observe the results shown in Fig.~\ref{fig:figure2} and Fig.~\ref{fig:figure4}, we use the following experimental parameters: a trap depth of $h \times 58$~MHz, oscillation frequencies $2\pi\times136$~kHz and $2\pi\times22$~kHz for the radial and axial dimensions, respectively, an atomic temperature of $\SI{107}{\micro\kelvin}$, and a bias magnetic field of 8.5 Gauss. For Fig.~\ref{fig:figure3}, we use a trap depth of $h \times10$~MHz, oscillation frequencies $2\pi\times56$~kHz and $2\pi\times9$~kHz for the radial and axial dimensions, respectively, and an atomic temperature of $\SI{44}{\micro\kelvin}$.

The detection of atoms in a particular $\left|m\right\rangle$ of the $f=2$ manifold is done by ejecting the atoms out of the trap. In the presence of the magnetic field, we use a Raman process to transfer only the population in the specific $\left|m\right\rangle$ to the $f=3$ manifold. We then deplete the $f=3$ population using the push out technique~\cite{Meschede2005} and then measure the number of remaining atoms in the trap using fluorescence detection~\cite{Andersen2015}. This procedure yields that the lost atoms were in the detected $\left|m\right\rangle$ while the remaining atoms were in the other states. In our push out technique, the detection efficiencies are $0.944\pm0.004$ and $0.997\pm0.003$ for the $f=2$ and $f=3$ states, respectively. In Fig.~\ref{fig:figure4} the probability for $\left|0,0 \right\rangle$ ($\hat{S}\left|1,-1 \right\rangle$) [$\hat{S}\left|2,-2 \right\rangle$] is determined by measuring the probability that zero atoms remain after atoms in the $\left|0\right\rangle$ ($\left|1\right\rangle$ and $\left|-1\right\rangle$) [$\left|2\right\rangle$ and $\left|-2\right\rangle$] are expelled. \\

\noindent \textbf{Relative number squeezing.}
The correlations between the $\left|\pm1\right\rangle$ of the two atoms (shown in Fig.~\ref{fig:figure2}) can be quantified by computing the population imbalance $J_z=\left(N_{+1}-N_{-1}\right)/2$, and the relative number squeezing~\cite{You2017} $\zeta^2=\frac{\left(\Delta J_z\right)^2}{N/4}$. $\Delta J_z$ is the standard deviation of $J_z$, $N_{\pm1}$ is number of atoms in $\left|\pm1\right\rangle$, and $N$ is the total number of atoms. We deduce the number squeezing from the data in Fig.~\ref{fig:figure2}c at the collision times of 150, 250, 350, and 500 ms (see Supplementary Note~3 for values of $\zeta^2$ at these individual collision times). If we postselect on at least one atom being detected in $\left|1\right\rangle$ or $\left|-1\right\rangle$, the result of ejecting atoms from both $\left|-1\right\rangle$ and $\left|1\right\rangle$ have only two possible outcomes: (1) zero atoms remain in the tweezer, which indicates that one atom was in $\left|-1\right\rangle$ and another was in $\left|1\right\rangle$, and therefore $J_z(n=0)=0$; or (2) one atom remains after ejection, which indicates that one atom was in $\left|\pm1\right\rangle$ and the other was in $\left|0\right\rangle$, $\left|-2\right\rangle$ or $\left|2\right\rangle$, so consequently $J_z(n=1)=\pm0.5$. Here, we assume that the probability of having both atoms in $\left|1\right\rangle$ or $\left|-1\right\rangle$ is zero. 

Still restricting to the subspace where at least one atom is in $\left|1\right\rangle$ or $\left|-1\right\rangle$ and taking $P_n$ to be the probability of $n$ atoms remaining in the optical tweezer after ejection, we can determine that the mean population imbalance is zero:
\begin{align}
	\left\langle J_z\right\rangle 	&=\frac{1}{\left(P_0+P_1\right)}\sum_{n=0,1} J_z\left(n\right)P_n \notag \\
							&=\frac{\left(0\times P_0+0.5\frac{P_1}{2}-0.5\frac{P_1}{2}\right)}{\left(P_0+P_1\right)}=0.
\end{align}
The variance of the population imbalance, $(\Delta J_z)^2 = \left\langle J_z^2\right\rangle - \left\langle J_z\right\rangle^2$, is given by:
\begin{align}
	(\Delta J_z)^2 	&= \left\langle J_z^2\right\rangle=\frac{1}{\left(P_0+P_1\right)}\sum_{n=0,1} \left(J_z\left(n\right)\right)^2P_n \notag \\
				&=\frac{0^2P_0+0.5^2P_1}{\left(P_0+P_1\right)}.
\end{align}
This allows us to quantify the degree of correlations between $|1\rangle$ and $|-1\rangle$ via the number squeezing parameter.
From above, the number squeezing is given by
\begin{equation}
	\zeta^2=\frac{\left(\Delta J_z\right)^2}{N/4}=\frac{P_1}{N \left( P_0+P_1 \right)}.
\end{equation}

Our measurement of the correlation can be influenced by the detection efficiency since the detection error in both $f=2$ and $f=3$ states will contribute to the measured value of $P_1$. The directly measured variance $\left(\Delta J_z\right)^2$ is $0.032\pm0.002$, while the detection error gives a variance of $0.034\pm0.002$ under the assumption that the actual $\left(\Delta J_z\right)^2=0$. This shows the measured degree of relative number squeezing can be entirely attributed to the detection efficiency. \\

\noindent \textbf{Coupling strengths and rate equations.}
We deduce the transition rates from the spin-dependent interaction strengths. We assume that the transition rates between $\hat{S}|m,-m\rangle$ and $\hat{S}|m',-m'\rangle$ are incoherent and have strengths proportional to $|\langle m',-m' | \hat{S} \hat{H}_\textrm{s} \hat{S}|m,-m\rangle|^2$. For low collisional energy, the interaction Hamiltonian of two atoms is approximated by~\cite{Bloch2005}
\begin{equation}
	\hat{H}_\textrm{s} = V\left(\hat{\textbf{r}}\right)\sum_{m_1, m_2, m'_1, m'_2}g_{m_1,m_2}^{m'_1,m'_2} \left|m'_1, m'_2 \right\rangle \left\langle m_1,m_2\right| \label{spin_exchange_Ham},
\end{equation}
where $\hat{\textbf{r}}$ is the relative position. The coupling coefficient between the initial $\left|m_1, m_2 \right\rangle$ and final $\left|m'_1, m'_2 \right\rangle$ of the atom pair is
\begin{equation}
g_{m_1,m_2}^{m'_1,m'_2}=\sum_{F=0}^{2f}\sum_{M=-F}^{F}g_{F}\left\langle m'_1,m'_2|F, M \right\rangle\left\langle F,M|m_1,m_2 \right\rangle,
\label{eq:2}
\end{equation}
where $g_F = 4 \pi \hbar^2 a_F / m$ with $a_F$ the $s$-wave scattering length for two atoms colliding in a channel with total spin $F$. As shown in Supplementary Note~4, provided both spin-2 atoms are initially prepared in the $m = 0$ Zeeman state, there are only six unique coupling coefficients in the above sum:
\begin{equation}
\begin{split}
g_{0,0}^{0,0}&=\frac{1}{35}\left(7g_0+10g_2+18g_4\right),\\
g_{0,0}^{1,-1}&=\frac{1}{35}\left(-7g_0-5g_2+12g_4\right),\\
g_{0,0}^{2,-2}&=\frac{1}{35}\left(7g_0-10g_2+3g_4\right),\\
g_{1,-1}^{1,-1}&=\frac{1}{70}\left(14g_0+5g_2+16g_4\right),\\
g_{1,-1}^{2,-2}&=\frac{1}{35}\left(-7g_0+5g_2+2g_4\right),\\
g_{2,-2}^{2,-2}&=\frac{1}{70}\left(14g_0+20g_2+g_4\right).
\end{split}
\label{eq:3}
\end{equation}
For $^{85}$Rb, the theoretically-predicted $s$-wave scattering lengths are $a_0=-740\pm60$ a.u., $a_2=-570\pm50$ a.u., and $a_4=-390\pm20$ a.u.~\cite{Greene2001}. By assuming the transition rate $\gamma_{mm'}$ between $\hat{S}\left|m,-m\right\rangle$ and $\hat{S}\left|m',-m'\right\rangle$ is proportional to $|\langle m',-m' | \hat{S} \hat{H}_\textrm{s} \hat{S}|m,-m\rangle|^2$, we get $\gamma_{01}/\gamma_{12}=\left(\sqrt{2}g_{0,0}^{1,-1}\right)^2/\left(2g_{1,-1}^{2,-2}\right)^2 = 2.34\pm1.66$. Similarly, $\gamma_{02}/\gamma_{01}$ and $\gamma_{02}/\gamma_{12}$ equal $0.04^{+0.08}_{-0.04}$ and $0.09^{+0.19}_{-0.09}$ respectively. We therefore set $\gamma_{02}$ to zero in the following rate equations.

Ignoring $\gamma_{02}$, we use the following rate equation to model the experimental results in Fig.~\ref{fig:figure4}:
\begin{align}
\frac{dP_{\left|0,0\right\rangle}}{dt}&=-\gamma_{01}P_{\left|0,0\right\rangle}+\gamma_{01}P_{\hat{S}\left|1,-1\right\rangle} \notag \\
\frac{dP_{\hat{S}\left|1,-1\right\rangle}}{dt}&=\gamma_{01}P_{\left|0,0\right\rangle}-\left(\gamma_{01}+\gamma_{12}\right)P_{\hat{S}\left|1,-1\right\rangle}+\gamma_{12}P_{\hat{S}\left|2,-2\right\rangle} \notag \\
\frac{dP_{\hat{S}\left|2,-2\right\rangle}}{dt}&=\gamma_{12}P_{\hat{S}\left|1,-1\right\rangle}-\gamma_{12}P_{\hat{S}\left|2,-2\right\rangle}
\label{eq:1}
\end{align}
where $P_{\hat{S}\left|m,-m\right\rangle}$ is the $\hat{S}\left|m,-m\right\rangle$ population. Using the above ratio of rates, we set $\gamma_{01}=2.34\times \gamma_{12}$ and fit the entire experimental dataset in Fig.~\ref{fig:figure4} using the single fitting parameter $\gamma_{12}$. \\

\noindent \textbf{Theoretical model of collisional spin dynamics.}
We describe the collisional dynamics of two bosonic atoms in a three-dimensional anisotropic harmonic potential with Hamiltonian Eq.~(\ref{Hamiltonian}) and spin-changing interaction given by Eq.~(\ref{spin_exchange_Ham}). As discussed above,
since both $F = 2$ atoms are initially prepared in the $m = 0$ Zeeman state, binary collisions preserve the spin projection along the quantization axis. Consequently, only three two-particle spin states are accessible: $\left|0,0\right\rangle$, $\hat{S}\left|1,-1\right\rangle$, and $\hat{S}\left|2,-2\right\rangle$. Writing the quantum state $|\psi(t)\rangle = \sum_{m=0,1,2} \int d\textbf{r} \, \psi_m(\textbf{r},t)| \textbf{r}\rangle \otimes \hat{S}|m,-m\rangle$, where $\hat{\textbf{r}} |\textbf{r} \rangle = \textbf{r} |\textbf{r}\rangle$, allows us to express the evolution under Hamiltonian (\ref{Hamiltonian}) as
\begin{align}
	i \hbar \dot{\psi}_0(\textbf{r})	&= H_\text{rel}(\textbf{r}) \psi_0(\textbf{r}) \notag \\
							&+ V(\textbf{r}) \left[ g_{0,0}^{0,0} \psi_0(\textbf{r}) + \sqrt{2} g_{0,0}^{1,-1} \psi_1(\textbf{r}) + \sqrt{2}  g_{0,0}^{2,-2} \psi_2(\textbf{r})\right], \notag \\
	i \hbar \dot{\psi}_1(\textbf{r})	&= \left(H_\text{rel}(\textbf{r}) + \hbar q_1 B^2 \right) \psi_1(\textbf{r}) \notag \\
							&+ V(\textbf{r}) \left[\sqrt{2} g_{0,0}^{1,-1} \psi_0(\textbf{r}) + 2 g_{1,-1}^{1,-1} \psi_1(\textbf{r}) + 2  g_{1,-1}^{2,-2} \psi_2(\textbf{r})\right], \notag \\
	i \hbar \dot{\psi}_2(\textbf{r})	&= \left(H_\text{rel}(\textbf{r}) + \hbar q_2 B^2 \right) \psi_2(\textbf{r}) \notag \\
							&+ V(\textbf{r}) \left[\sqrt{2} g_{0,0}^{2,-2} \psi_0(\textbf{r}) + 2 g_{1,-1}^{2,-2} \psi_1(\textbf{r}) + 2  g_{2,-2}^{2,-2} \psi_2(\textbf{r})\right],
\label{coupled_eqs}
\end{align}
where $H_\text{rel}(\textbf{r}) = -\frac{\hbar^2}{2\mu}\nabla^2_\textbf{r} + \frac{1}{2}\sum_{i=x,y,z}\mu\omega_i^2 r_i^2$, the coupling constants are given by Eq.~(\ref{eq:3}), and the quadratic Zeeman shifts are $q_1 = 143.776$ Hz/G$^2$ and $q_2 = 575.104$ Hz/G$^2$~~\cite{Run-Bing:2009}.

We take our initial condition as $\psi_1(\textbf{r},0) = \psi_2(\textbf{r},0) = 0$ and $\psi_0(\textbf{r},0)$ as a thermal distribution of even eigenstates of $H_\text{rel}(\textbf{r})$. Specifically, in any given experiment $\psi_0(\textbf{r},0) = \varphi_{n_x}(x)\varphi_{n_y}(y)\varphi_{n_z}(z)$, where $\varphi_{n_i}(x_i)$ are eigenstates of the 1D harmonic oscillator with mass $\mu$ and frequency $\omega_i$ and $(-1)^{n_x+n_y+n_z} = 1$ (since $\psi_0(\textbf{r})$ must be symmetric under particle exchange). The Boltzmann probability that $\psi_0(\textbf{r}, 0)$ will be prepared in the eigenstate with quantum numbers $(n_x, n_y, n_z)$ is $\mathcal{P}(n_x,n_y,n_z) = \exp(-\beta \varepsilon_{n_x,n_y,n_z}) / \mathcal{Z}$,
where $\varepsilon_{n_x,n_y,n_z} = \hbar \omega_x (n_x+\tfrac{1}{2})+\hbar \omega_y (n_y+\tfrac{1}{2})+\hbar \omega_z (n_z+\tfrac{1}{2})$ are the eigenstate energies, $\beta = 1 / k_\textrm{B} T$, and the partition function $\mathcal{Z}$ has an analytic expression (see Supplementary Note~5).

In the low-energy regime where $s$-wave collisions dominate, it is customary to take $V(\textbf{r}) = \delta(\textbf{r})$~\cite{Bloch2005}. However, in this case spin-changing dynamics only occur for eigenstates where $n_x, n_y, n_z$ are all even (see Supplementary Note~6). In contrast, states where (say) $n_x$ is even and $n_y$ and $n_z$ are odd never evolve. These latter kinds of states represent roughly $70\%$ of the ensemble at $44~\mu$K, implying that this model predicts that the population of $|0,0\rangle$ never drops below $0.7$, at odds with what we experimentally observe.

We wish to use a simplified atom-atom interaction model that allows for numerical calculations involving a high number of modes, while at the same time avoids the problem with the delta-function interaction model~\cite{Blume:2012}. In particular, there is some evidence that the zero-range $\delta$-function pseudopotential fails to replicate the scattering properties of the underlying physical potential in trapped systems when the magnitude of the $s$-wave scattering length is on the order or greater than the harmonic oscillator lengthscale~\cite{Tiesinga:2000, Blume:2002}. Further, there is a greater discrepancy for negative scattering lengths. In our experiment $a_0 / d = -0.44$, $a_2 / d = -0.34$, and $a_4/d = -0.23$, where $d = \sqrt{\hbar / (m \bar{\omega})}$ and $\bar \omega = (\omega_x \omega_y \omega_z)^{1/3}$.
We use a Gaussian pseudopotential $V(\textbf{r}) = \exp[-r^2 / (2w^2)]/(2\pi w^2)^{3/2}$ with $w = \sqrt{(a_0^4 + a_2^4 + a_4^4) / (a_0^2 + a_2^2 + a_4^2)} \sim 650$~a.u., since (1) it is finite range and couples all even-parity eigenstates, (2) it gives the same total scattering cross section as the $\delta$-function pseudopotential (see Supplementary Note~7), (3) it smoothly recovers the (regularised) $\delta$-function in the $w \to 0$ limit, and (4) the form of the spin-changing coupling matrix is sufficiently simple that a numeric calculation is tractable.

We numerically solve for the spin-changing dynamics by expanding $\psi_i(\textbf{r})$ on a finite basis of even-parity eigenstates of $H_\text{rel}(\textbf{r})$: $\psi_i(\textbf{r},t) = \sum_{\varepsilon_{n_x,n_y,n_z} \leq E_\text{cut}} c_{n_x,n_y,n_z}^i(t) \varphi_{n_x}(x)\varphi_{n_y}(y)\varphi_{n_z}(z)$, where the sum is over all eigenstates with energy $\epsilon_{n_x,n_y,n_z}$ less than some energy cutoff $E_\text{cut}$. It is necessary to choose $E_\text{cut}$ sufficiently large that $\sum_{\varepsilon_{n_x,n_y,n_z} \leq E_\text{cut}}\mathcal{P}(n_x,n_y,n_z) \approx 1$ and coupling to the highest-energy, sparsely-occupied modes is negligible. For the computational resources at our disposal, these conditions limit our calculations to temperatures no greater than $\SI{8.8}{\micro \kelvin}$ -- roughly one fifth the temperature of the experiment.

In this basis the state is represented by $\textbf{c} = [\textbf{c}^0, \textbf{c}^1, \textbf{c}^2]^\top$, where $\textbf{c}^i$ is the vector of coefficients $c_{n_x,n_y,n_z}^{i}$ for modes satisfying $\epsilon_{n_x,n_y,n_z} \leq E_\text{cut}$. Equations~(\ref{coupled_eqs}) imply $i \hbar \dot{\textbf{c}}(t) = \textbf{H} \, \textbf{c}(t)$ with
\begin{widetext}
\begin{equation}
	\textbf{H} = 	\begin{pmatrix}
					\bm{\epsilon} +  g_{0,0}^{0,0} \textbf{T}	&& \sqrt{2}  g_{0,0}^{1,-1} \textbf{T}		&& \sqrt{2}  g_{0,0}^{2,-2} \textbf{T} \\
					\sqrt{2}  g_{0,0}^{2,-2} \textbf{T}	&& (\bm{\epsilon} + \hbar q_1 B^2 \textbf{I}) + 2 g_{1,-1}^{1,-1} \textbf{T}		&& 2 g_{1,-1}^{2,-2} \textbf{T} \\
					\sqrt{2}  g_{0,0}^{2,-2} \textbf{T}	&& 2 g_{1,1}^{2,-2} \textbf{T}		&& (\bm{\epsilon} + \hbar q_1 B^2 \textbf{I}) + 2 g_{2,-2}^{2,-2} \textbf{T}
				\end{pmatrix}. \label{Ham_matrix}
\end{equation}
\end{widetext}
Here $\bm{\epsilon}$ is a diagonal matrix with energies $\varepsilon_{n_x,n_y,n_z}$ along the diagonal and the coupling matrix $\textbf{T}$ is defined via $T_{n_x,n_y,n_z}^{m_x,m_y,m_z} = \mathcal{I}_{n_x,m_x} \mathcal{I}_{n_y,m_y} \mathcal{I}_{n_z,m_z} / (2\pi w^2)^{3/2}$, where the integrals $\mathcal{I}_{n_i,m_i} = \int dx_i \, \varphi_{n_i}(x_i) \exp[-x_i^2 / (2 w^2)] \varphi_{m_i}^i(x_i)$ have an analytic solution in terms of Gauss hypergeometric functions (see Supplementary Note~8). Diagonalising $\textbf{H} = \textbf{U} \textbf{D} \textbf{U}^\dag$ gives the solution $\textbf{c}(t) = \textbf{U} \exp[-\tfrac{i}{\hbar}\textbf{D} t] \textbf{U}^\dag \textbf{c}(0)$. 
Thus, for a given initial condition $\psi_0(\textbf{r},0) = \varphi_{m_x}(x)\varphi_{m_y}(y)\varphi_{m_z}(z)$ we can compute the population of the $j$th two-boson spin state $N_{m_x,m_y,m_z}^j(t) = \sum_{\varepsilon_{n_x,n_y,n_z} \leq E_\text{cut}} |c_{n_x,n_y,n_z}^j(t)|^2$. 
The total population of the $j$th two-boson spin state assuming a thermal initial state is given by an incoherent sum over $N_{m_x,m_y,m_z}^j(t)$ weighted by the Boltzmann probability $\mathcal{P}(m_x,m_y, m_z)$:
\begin{equation}
	P_{|j,-j\rangle}(t) = \sum_{\varepsilon_{m_x,m_y,m_z} \leq E_\text{cut}} \mathcal{P}(m_x,m_y, m_z) N_{m_x,m_y,m_z}^j(t).
\end{equation}
This procedure was used to generate the simulation data plotted in Fig.~\ref{fig:figure3}, Fig.~\ref{fig:figure4}, and Fig.~\ref{fig:figure5}.

\section*{Data and code availability}
The source data underlying Figs 2 to 5 and Supplementary Figure~1 are provided as a Source Data file. Simulation codes are available from Stuart S. Szigeti on reasonable request.

\section*{AUTHOR CONTRIBUTIONS}
The experiments were carried out by P.S. and E.S. The data analysis was performed by P.S. Theoretical and numerical analysis was performed by S.S.S. and supervised by A.S.B. All work was supervised by M.F.A.. All authors discussed the results and contributed to the manuscript.

\bibliographystyle{apsrev4-1}

%\bibliographystyle{naturemag}
%\bibliography{sample}

\begin{thebibliography}{1}
 
\bibitem{Horvath2017} Horvath, M., Thomas, R., Tiesinga, E., Deb, A., \& Kjaergaard, N. Above-threshold scattering about a Feshbach resonance for ultracold atoms in an optical collider. \textit{Nat. Comm.} \textbf{8}, 452 (2017).
 
\bibitem{Isenhower2010} Isenhower, L., Urban, E., Zhang, X.~L., Gill, A.~T., Henage, T., Johnson, T.~A., Walker, T.~G., \& Saffman, M. Demonstration of a neutral atom controlled-NOT quantum gate. \textit{Phys. Rev. Lett.} \textbf{104}, 010503 (2010).

\bibitem{Bloch2012} Bloch, I., Dalibard, J., \& Nascimbene, S. Quantum simulations with ultracold quantum gases. \textit{Nat. Phys.} \textbf{8}, 267 -- 276 (2012).

\bibitem{Greiner2016} Kaufman, A.~M., Tai, M.~E., Lukin, A., Rispoli, M., Schittko, R., Preiss, P.~M., \& Greiner, M. Quantum thermalization through entanglement in an isolated many-body system. \textit{Science} \textbf{353}, 794 -- 800 (2016).

\bibitem{Lahav2010} Lahav, O., Itah, A., Blumkin, A., Gordon, C., Rinott, S., Zayats, A., \& Steinhauer, J. Realization of a Sonic Black Hole Analog in a Bose-Einstein Condensate. \textit{Phys. Rev. Lett.} \textbf{105} 240401 (2010).

\bibitem{Greiner2003} Greiner, M., Regal, C.~A. \& Jin D.~S. Emergence of a molecular Bose-Einstein condensate from a Fermi gas. \textit{Nature} \textbf{426}, 537 -- 540 (2003).

\bibitem{Williamson2017} Williamson, L. A., \& Blakie, P. B. Coarsening Dynamics of an Isotropic Ferromagnetic Superfluid. \textit{Phys. Rev. Lett.} \textbf{119}, 255301 (2017).

\bibitem{Stamper-Kurn2016} Fang, F., Olf, R., Wu, S., Kadau, H.,\& Stamper-Kurn, D. M. Condensing magnons in a degenerate ferromagnetic spinor Bose gas. \textit{Phys. Rev. Lett.} \textbf{116}, 095301 (2016).

\bibitem{Sengstock2004} Schmaljohann, H., Erhard, M., Kronj\"ager, J., Kottke, M., Van Staa, S., Cacciapuoti, L., Arlt, J.J., Bongs, K. \& Sengstock, K. Dynamics of $F=2$ spinor Bose-Einstein condensates. \textit{Phys. Rev. Lett.} \textbf{92}, 040402 (2004).

\bibitem{Chapman2005} Chang, M.~S., Qin, Q., Zhang, W., You, L., \& Chapman, M.~S. Coherent spinor dynamics in a spin-1 Bose condensate. \textit{Nature} \textbf{1}, 111 - 116 (2005).

\bibitem{Lett2013} Pechkis, H.~K., Wrubel, J.~P., Schwettmann, A., Griffin, P.~F., Barnett, R., Tiesinga, E., \& Lett, P.~D. Spinor dynamics in an antiferromagnetic spin-1 thermal Bose gas. \textit{Phys. Rev. Lett.} \textbf{111}, 025301 (2013).

\bibitem{Ebling2014} Ebling, U., Krauser, J.~S., Fl{\"a}schner, N., Sengstock, K., Becker, C., Lewenstein, M., \& Eckardt, A. Relaxation Dynamics of an Isolated Large-Spin Fermi Gas Far from Equilibrium. \textit{Phys. Rev. X} \textbf{4}, 021011 (2014). 

\bibitem{Krauser:2014} Krauser, J.~S., Ebling, U., Fl{\"a}schner, N., Heinze, J., Sengstock, K., Lewenstein, M., Eckardt, A., \& Becker, C. Giant Spin Oscillations in an Ultracold Fermi Sea. \textit{Science} \textbf{343}, 157 -- 160 (2014). 

\bibitem{He:2015} He, X., Zhu, B., Li, X., Wang, F., Xu, Z.~F. \& Wang, D. Coherent spin-mixing dynamics in thermal $^{87}\mathrm{Rb}$ spin-1 and spin-2 gases. \textit{Phys. Rev. A} \textbf{91},033635 (2015).

\bibitem{Hoang:2016} Hoang, T.~M., Anquez, M., Robbins, B.~A., Yang, X.~Y., Land, B.~J., Hamley, C.~D., \& Chapman, M.~S. Parametric excitation and squeezing in a many-body spinor condensate. \textit{Nat. Commun.} \textbf{7}, 11233 (2016).

\bibitem{You2017} Luo, X.~Y., Zou, Y.~Q., Wu, L.~N., Liu, Q., Han, M.~F., Tey, M.~K., \& You, L. Deterministic entanglement generation from driving through quantum phase transitions. \textit{Science} \textbf{355}, 620 -- 623 (2017).

\bibitem{Klempt2011} L{\"u}cke, B., Scherer, M., Kruse, J., Pezz\'e, L., Deuretzbacher, F., Hyllus, P., Topic, O., Peise, J., Ertmer, W., Arlt, J., Santos, L., Smerzi, A., \& Klempt, C. Twin matter waves for interferometry beyond the classical limit. \textit{Science} \textbf{334}, 773 -- 776 (2011).

\bibitem{Linnemann:2016} Linnemann, D., Strobel, H., Muessel, W., Schulz, J., Lewis-Swan, R.~J., Kheruntsyan, K.~V., \& Oberthaler, M.~K. Quantum-Enhanced Sensing Based on Time Reversal of Nonlinear Dynamics. \textit{Phys. Rev. Lett.} \textbf{117} 013001 (2016).

\bibitem{Fadel:2018} Fadel, M., Zibold, T., D\'ecamps, B., \& Treutlein, P. Spatial entanglement patterns and Einstein-Podolsky-Rosen steering in Bose-Einstein condensates. \textit{Science}, \textbf{360}, 409-413 (2018). 

\bibitem{Kunkel2018} Kunkel, P., Pr{\"u}fer, M., Strobel, H., Linnemann, D., Fr{\"o}lian, A., Gasenzer, T., G{\"a}rttner, M., \& Oberthaler, M.~K. Spatially distributed multipartite entanglement enables EPR steering of atomic clouds. \textit{Science}, \textbf{360}, 413 -- 416 (2018). 

\bibitem{Lange:2018} Lange, K., Peise, J., L\"ucke, B., Kruse, I., Vitagliano, G., Apellaniz, I., Kleinmann, M., T\'oth, G., \& Klempt, C. Entanglement between two spatially separated atomic modes. \textit{Science}, \textbf{360}, 416 -- 418 (2018).

\bibitem{Grimm2006} Kraemer, T., Mark, M., Waldburger, P., Danzl, J.~G., Chin, C., Engeser, B., Lange, A.~D., Pilch, K., Jaakkola, A., N\"agerl, H.~-C., \& Grimm, R. Evidence for Efimov quantum states in an ultracold gas of caesium atoms. \textit{Nature} \textbf{440}, 315 -- 318 (2006).

\bibitem{Salomon2017} Laurent, S., Pierce, M., Delehaye, M., Yefsah, T., Chevy, F., \& Salomon, C.  Connecting Few-Body Inelastic Decay to Quantum Correlations in a Many-Body System: A Weakly Coupled Impurity in a Resonant Fermi Gas. \textit{Phys. Rev. Lett.} \textbf{118}, 103403 (2017).

\bibitem{Bloch2005} Widera, A., Gerbier, F., F\"olling, S., Gericke, T., Mandel, O., \& Bloch, I. Coherent collisional spin dynamics in optical lattices. \textit{Phys. Rev. Lett.} \textbf{95}, 190405 (2005). 

\bibitem{Widera:2006} Widera, A., Gerbier, F., F\"olling, S., Gericke, T., Mandel, O., \& Bloch, I. Precision measurement of spin-dependent interaction strengths for spin-1 and spin-2 87Rb atoms. \textit{New J. Phys.} \textbf{8}, 152 (2006).

\bibitem{Greene2001} Klausen, N.~N., Bohn, J.~L., \& Greene, C.~H. Nature of spinor Bose-Einstein condensates in rubidium. \textit{Phys. Rev. A} \textbf{64}, 053602 (2001).

\bibitem{Endres2016} Endres, M., Bernien, H., Keesling, A., Levine, H., Anschuetz, E.~R., Krajenbrink, A., Senko, C., Vuletic, V., Greiner, M., \& Lukin, M.~D. Atom-by-atom assembly of defect-free one-dimensional cold atom arrays. \textit{Science} \textbf{354}, 1024 -- 1027 (2016).

\bibitem{Barredo:2016} Barredo, D., de~L{\'e}s{\'e}leuc, S., Lienhard, V., Lahaye, T., \& Browaeys, A. An atom-by-atom assembler of defect-free arbitrary 2D atomic arrays. \emph{Science} \textbf{354}, 1021 -- 1023 (2016).

\bibitem{Andersen2010} Gr{\"u}nzweig, T., Hilliard, A., McGovern, M., \& Andersen, M.~F. Near-deterministic preparation of a single atom in an optical microtrap. \textit{Nat. Phys.} \textbf{6}, 951 -- 954 (2010). 

\bibitem{Andersen2013} Sompet, P., Carpentier, A.~V., Fung, Y.~H., McGovern, M., \& Andersen, M.~F. Dynamics of two atoms undergoing light-assisted collisions in an optical microtrap. \textit{Phys. Rev. A} \textbf{88}, 051401 (2013).

\bibitem{Liu2018} Liu, L.~R., Hood, J.~D., Yu, Y., Zhang, J.~T., Hutzler, N.~R., Rosenband, T., \& Ni, K.-K. Building one molecule from a reservoir of two atoms. \textit{Science}, \textbf{360}, 900 -- 903 (2018).

\bibitem{Xu2015} Xu, P., Yang, J., Liu, M. He, X., Zeng, Y., Wang, K., Wang, J., Papoular, D.~J., Shlyapnikov, G.~V., \& Zhan, M. Interaction-induced decay of a heteronuclear two-atom system, \textit{Nat. Commun.} \textbf{6}, 780, (2015).

\bibitem{Regal2015} Kaufman, A. M., Lester, B. J., Foss-Feig, M., Wall, M. L., Rey, A. M., \& Regal, C. A. Entangling two transportable neutral atoms via local spin exchange. \textit{Nature} \textbf{527}, 208--211 (2015).

%\bibitem{Andersen2016} Fung, Y.~H., Sompet, P., \& Andersen, M.~F. Single Atoms Preparation Using Light-Assisted Collisions. \textit{Technologies} \textbf{4}, 4 (2016).
\bibitem{Andersen2013_2} Carpentier, A.~V., Fung, Y.~H., Sompet, P., Hilliard, A.~J., Walker, T.~G., \& Andersen, M.~F. Preparation of a single atom in an optical microtrap. \textit{Laser Phys. Lett.} \textbf{10},12 (2013).

\bibitem{Regal2015a} Lester, B.~J., Luick, N., Kaufman, A.~M., Reynolds, C.~M., \& Regal, C.~A. Rapid production of uniformly filled arrays of neutral atoms. \textit{Phys. Rev. Lett.}, \textbf{115,} 073003 (2015).

\bibitem{Tiesinga:2000} Tiesinga, E., Williams, C.~J., Mies, F.~H., \& Julienne, P.~S. Interacting atoms under strong quantum confinement. \emph{Phys. Rev. A} \textbf{61}, 063416 (2000).

\bibitem{Blume:2002} Blume, D. \& Greene, C.~H. Fermi pseudopotential approximation: Two particles under external confinement. \emph{Phys. Rev. A} \textbf{65}, 043613 (2002).

\bibitem{Stamper-Kurn2013} Stamper-Kurn, D.~M., \& Ueda, M. Spinor Bose gases: Symmetries, magnetism, and quantum dynamics. \textit{Rev. Mod. Phys.} \textbf{85}, 1191 (2013).

\bibitem{Klempt2014} L\"ucke, B., Peise, J., Vitagliano, G., Arlt, J., Santos, L., T\'oth, G., \& Klempt, C. Detecting multiparticle entanglement of Dicke states. \textit{Phys. Rev. Lett.} \textbf{112}, 155304 (2014).

\bibitem{Oktel:2002} Oktel, M. \"O. and Levitov, L. S. Internal Waves and Synchronized Precession in a Cold Vapor. \textit{Phys. Rev. Lett.} \textbf{88} 230403 (2002). 

\bibitem{Williams:2002} Willams, J. E., Nikuni, T., Clark, C. W. Longitudinal Spin Waves in a Dilute Bose Gas. \textit{Phys. Rev. Lett.} \textbf{88} 230405 (2002). 

\bibitem{Fuchs:2002} Fuchs, J. N., Gangardt, D. M., Lalo\"e, F. Internal State Conversion in Ultracold Gases \textit{Phys. Rev. Lett.} \textbf{88} 230404 (2002). 

\bibitem{Bradley:2002} Bradley, A S and Gardiner, C W Theory of Ramsey spectroscopy and anomalous segregation in ultracold rubidium. \textit{J. Phys. B} \textbf{20}, 4299 -- 4323 (2002).

%\bibitem{Grimm2010} Chin, C., Grimm, R., Julienne, P. \& Tiesinga, E. Feshbach resonances in ultracold gases. \textit{Rev. Mod. Phys.}, \textbf{82},1225 (2010). \hl{I do not think this paper discuss what we cite it for!}
\bibitem{Thompson2013} Thompson, J. D., Tiecke, T. G., Zibrov, A. S., Vuletic, V. \& Lukin, M. D. Coherence and Raman Sideband Cooling of a Single Atom in an Optical Tweezer. \textit{Phys. Rev. Lett.} \textbf{110}, 133001 (2013).

\bibitem{Blume:2012} Blume, D. Few-body physics with ultracold atomic and molecular systems in traps. \emph{Reports on Progress in Physics} \textbf{75}, 046401 (2012).

\bibitem{Meschede2005} Kuhr, S., Alt, W., Schrader, D., Dotsenko, I., Miroshnychenko, Y., Rauschenbeutel, A., \& Meschede, D. Analysis of dephasing mechanisms in a standing-wave dipole trap. \textit{Phys. Rev. A} \textbf{72}, 023406 (2005).

\bibitem{Andersen2015} Hilliard, A.J., Fung, Y.H., Sompet, P., Carpentier, A.V., \& Andersen, M.F. In-trap fluorescence detection of atoms in a microscopic dipole trap. \textit{Phys. Rev A} \textbf{91}, 053414 (2015).

\bibitem{Pezze:2009} Pezz\'e, L. and Smerzi, A. Entanglement, Nonlinear Dynamics, and the Heisenberg Limit. \textit{Phys. Rev. Lett.} \textbf{102}, 100401 (2009).

\bibitem{Nielsen:2010} Nielsen, M. and Chuang, I.~L. \emph{Quantum Computation and Quantum Information: 10th Anniversaty Edition}. Cambridge University Press, Cambridge UK (2010).

%\bibitem{Regal2015b}Lester, Brian J., et al. Rapid production of uniformly filled arrays of neutral atoms. \textit{Phys. Rev. Lett.} \textbf{115,} 073003 (2015).
\bibitem{Run-Bing:2009} Run-Bing Li and Lin Zhou and Jin Wang and Ming-Sheng Zhan. Measurement of the quadratic Zeeman shift of 85Rb hyperfine sublevels using stimulated Raman transitions. \textit{Optics Communications} \textbf{7}, 1340 -- 1344 (2009).

%\bibitem{Toschek1986} Sauter, T., Neuhauser, W., Blatt, R., \& Toschek, P. E. Observation of quantum jumps. \textit{Phys. Rev. Lett.} \textbf{57,} 1696 (1986).

%\bibitem{Rosnagel:2016} Ro\ss nagel, J., Dawkins, S.~T., Tolazzi, K.~N., Abah,~O., Lutz, E., Schmidt-Kaler, F., \& Singer, K. A single-atom heat engine. \textit{Science} \textbf{352}, 325-329 (2016).

\end{thebibliography}

\begin{thebibliography}{1}
	\bibitem{Dalibard:1998} Dalibard, J. \emph{Collisional dynamics of ultracold atomic gases}, in Bose-Einstein Condensation in Atomic Gases 1998 321 - 349, (Course CXL of Proceedings of the International School of Physics ``Enrico Fermi'', IOS Press, 1998).
	\bibitem{Pezze:2009} Pezz\'e L. \& Smerzi, A. Entanglement, Nonlinear Dynamics, and the Heisenberg Limit. \emph{Phys. Rev. Lett.} \textbf{102}, 100401 (2009).
	\bibitem{Gradshteyn:2014} Gradshteyn I.~S. \& Ryzhik, I.~M. \emph{Table of Integrals, Series, and Products, 8th edition}. Academic Press (2014).
\end{thebibliography}

%% Here is the endmatter stuff: Supplementary Info, etc.
%% Use \item's to separate, default label is "Acknowledgements"

\section{Acknowledgements and additional information}
We acknowledge assistance with data acqusition by Tarentaise L. McLeod and fruitful discussions with Crispin~W.~Gardiner, Ryan Thomas, and the members of Blair Blakie's research group. This work was supported by the Marsden Fund Council from Government funding, administered by the Royal Society of New Zealand (Contract No. UOO1320) and the Dodd-Walls Centre for Photonic and Quantum Technologies. S.~S.~S received funding from an Australia Awards-Endeavour Research Fellowship and the Australian Research Council (Projects No.~DP160104965 and No.~DP150100356).

The authors declare that they have no competing interests.

Correspondence and requests for materials should be addressed to M.F.A.~(email: mikkel.andersen@otago.ac.nz).

\newpage \newpage

\setcounter{equation}{0}
\setcounter{figure}{0}

\begin{widetext}
\section{Supplementary Information}

In this Supplementary Information we provide further details on (1) the $\left|\pm2\right\rangle$ population dynamics, (2) the argument for the presence of thermally-robust, metrologically-useful entanglement in our atom-pair experiment, (3) the correlation between $|\pm 1\rangle$ at different collision times, (4) the spin-changing collisional coupling constants, (5) the analytic form for the initial thermal distribution used in our theoretical modelling, (6) why a $\delta$-function pseudopotential is incapable of modelling the dynamics of our experiment, (7) the justification of the width chosen for our Gaussian pseudopotential, and (8) the coupling matrix that arises from a Gaussian pseudopotential.

\subsection{Supplementary Note 1: $\left|\pm2\right\rangle$ population dynamics}
In addition to the $m$-state correlation results that we show in Fig.~2 of the main text, for $\left|0\right\rangle$ and $\left|\pm1\right\rangle$, we do the same measurement for the case of $\left|\pm2\right\rangle$. The results are shown in Supplementary Figure~\ref{fig:S1}.
Experimentally, the atom pair are initially prepared in $\left|0\right\rangle$. In Supplementary Figure~\ref{fig:S1}a, after a given collision time, we eject atoms in $\left|-2\right\rangle$ from the trap. Since the probability of one atom remaining in the trap increases with the collision time, this indicates that the population of $\left|-2\right\rangle$ increases with collision time. In the case of ejecting both atoms from $\left|-2\right\rangle$ and $\left|+2\right\rangle$, only pair loss is observed, as shown in Supplementary Figure~\ref{fig:S1}b. Combining Supplementary Figure~\ref{fig:S1}a with Supplementary Figure~\ref{fig:S1}b allows us to conclude that when one atom is in $\left|-2\right\rangle$, the other is in $\left|+2\right\rangle$. Therefore, the collisional dynamics result in correlations between the $\left|\pm2\right\rangle$ populations. However, the correlations between the $\left|\pm2\right\rangle$ populations are moderate compared to the measured $\left|\pm1\right\rangle$ case presented in Fig.~2b and Fig.~2c in the main text. This could be due to imperfect $\pi$-pulse Raman transfer for the detection of atoms in $\left|\pm2\right\rangle$ in the $F=2$ manifold, which is more sensitive to magnetic noise compared to the $\left|\pm1\right\rangle$ case.

%%%%%%%%%%%   Figure S1   %%%%%%%%%%%%
\begin{figure}
	\begin{center}
      \includegraphics[width=0.6\linewidth]{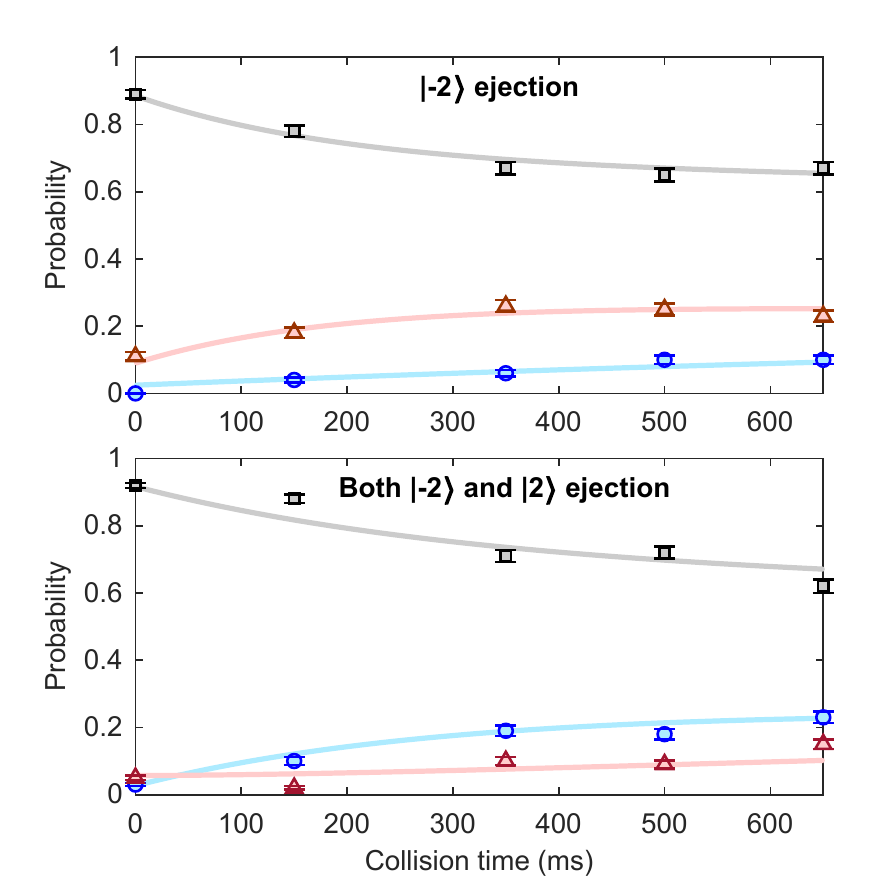}
	\end{center}
   	\caption{
	{\bf \boldsymbol{$\left|\pm2\right\rangle$ } population results.}
	 Probability that zero, one, or two atoms remain in the optical tweezer after a given collision time. The error bars represent the standard error of the mean. \textbf{a,} When atoms solely from $\left|-2\right\rangle$ are expelled (immediately after a given collision time), this gives only single-atom loss events, which is opposite to the result in \textbf{b}. \textbf{b,} Expelling atoms from both $\left|-2\right\rangle$ and $\left|2\right\rangle$ gives only pair loss. In all cases and throughout the collision time, the bias magnetic field was 8.5 Gauss. The solid curves are a fit to the measured data, used to guide the eye. Source data are provided as a Source Data file.}
	\label{fig:S1}
\end{figure}
%%%%%%%%%%%%%%%%%%%%%%%%%%%%%%

\subsection{Supplementary Note 2: Prospects of metrologically-useful entanglement generation}
Here we provide a detailed theoretical argument showing that, in principle, our experiment is capable of generating thermally-robust entangled states which are metrologically useful.

Suppose we prepare a thermal ensemble of atom pairs in $|0,0\rangle$ - i.e. $\hat{\rho}_0 = \left(\sum_\textbf{n} \mathcal{P}(\textbf{n})|\textbf{n}\rangle \langle \textbf{n}|\right) \otimes |0,0\rangle \langle 0,0| $, where $|\textbf{n}\rangle \equiv |n_x, n_y, n_z \rangle$ denotes the even-parity eigenstates of the harmonic oscillator Hamiltonian and $\mathcal{P}(\textbf{n})$ is a Boltzmann distribution. Then under Hamiltonian~(1) of the main text this initial state will always evolve to a state of the form
\begin{equation}
	\hat{\rho}(t) = \sum_\textbf{n} \mathcal{P}(\textbf{n}) |\psi_\textbf{n}(t)\rangle \langle \psi_\textbf{n}(t)|, \label{density_matrix}
\end{equation}
where
\begin{equation}
	 |\psi_\textbf{n}(t)\rangle = \sum_\textbf{m} \left[ c_\textbf{m}^0(\textbf{n},t)|0,0\rangle + c_\textbf{m}^1(\textbf{n},t)\hat{\mathcal{S}}|1,-1\rangle  + c_\textbf{m}^2(\textbf{n},t)\hat{\mathcal{S}}|2,-2\rangle \right]
\end{equation}
is the state that results from evolving the \emph{pure} initial state $|\textbf{n}\rangle$ under Hamiltonian~(1) of the main text. The coefficients $c_\textbf{m}^j(\textbf{n},t)$ are determined by the numerical procedure described in the Methods (for example, by diagonalising the matrix defined in Eq.~(10) of the Methods).

We now show that tracing out the motional degrees of freedom of Eq.~(\ref{density_matrix}) results in a reduced density matrix with off-diagonal elements indicative of entanglement between two-particle spin states $|1,-1\rangle$ and $|-1,1\rangle$ (and also between $|2,-2\rangle$ and $|-2,2\rangle$). Explicitly, the reduced density matrix that only accounts for the spin degrees of freedom is
\begin{align}
	\hat{\rho}_\textrm{S}(t)	&= \textrm{Tr}_\textrm{M}\left\{ \hat{\rho}(t) \right\} = \sum_\textbf{n} \mathcal{P}(\textbf{n}) \sum_\textbf{m} \langle \textbf{m}| \psi_\textbf{n}(t)\rangle \langle \psi_\textbf{n}(t) | \textbf{m} \rangle,
\end{align}
where $\textrm{Tr}_\textrm{M}$ denotes the partial trace over the motional degrees of freedom. Noting that
\begin{equation}
	\langle \textbf{m}|\psi_\textbf{n}(t)\rangle =  c_\textbf{m}^0(\textbf{n},t)|0,0\rangle + c_\textbf{m}^1(\textbf{n},t)\hat{\mathcal{S}}|1,-1\rangle  + c_\textbf{m}^2(\textbf{n},t) \hat{\mathcal{S}}|2,-2\rangle,
\end{equation}
and defining the coefficients
\begin{equation}
	\rho_{i,j}(t)	\equiv \sum_\textbf{n} \mathcal{P}(\textbf{n}) \sum_\textbf{m} c_\textbf{m}^i(\textbf{n},t) [c_\textbf{m}^j(\textbf{n},t)]^*,
\end{equation}
where $i,j = 0,1,2$, we can write $\hat{\rho}_\textrm{S}(t)$ as
\begin{align}
	\hat{\rho}_\textrm{S}(t)	&= \rho_{0,0}(t) |0,0\rangle \langle 0,0| + \tfrac{1}{2}\rho_{1,1}(t) \Big(|-1,1\rangle \langle -1,1| + |-1,1\rangle \langle 1,-1| + |1,-1\rangle \langle -1,1| + |1,-1\rangle \langle 1,-1| \Big) \notag \\
				&+ \tfrac{1}{2}\rho_{2,2}(t) \Big(|-2,2\rangle \langle -2,2| + |-2,2\rangle \langle 2,-2| + |2,-2\rangle \langle -2,2| + |2,-2\rangle \langle 2,-2| \Big) \notag \\
				&+ \Bigg[ \tfrac{1}{\sqrt{2}}\rho_{0,1}(t)\Big( |0,0\rangle \langle -1,1| + |0,0\rangle \langle 1,-1|\Big) + \tfrac{1}{\sqrt{2}}\rho_{0,2}(t)\Big( |0,0\rangle \langle -2,2| + |0,0\rangle \langle 2,-2|\Big) \notag \\
				&\qquad + \tfrac{1}{2}\rho_{1,2}(t)\Big( |-1,1\rangle \langle -2,2| + |-1,1\rangle \langle 2,-2| + |1,-1\rangle \langle 2,-2| + |1,-1\rangle \langle -2,2| \Big) + h.c. \Bigg],
\end{align}
or in matrix notation as
\begin{align}
			\hat{\rho}_\textrm{S}(t) &\equiv \begin{pmatrix}
					\rho_{0,0}(t)				&& \tfrac{1}{\sqrt{2}}\rho_{0,1}(t)		&& \tfrac{1}{\sqrt{2}}\rho_{0,1}(t)	&& \tfrac{1}{\sqrt{2}}\rho_{0,2}(t)	&& \tfrac{1}{\sqrt{2}}\rho_{0,2}(t) \\
					\tfrac{1}{\sqrt{2}}\rho_{0,1}^*(t)	&& \tfrac{1}{2}\rho_{1,1}(t)				&& \tfrac{1}{2}\rho_{1,1}(t)			&& \tfrac{1}{2}\rho_{1,2}(t)		&& \tfrac{1}{2}\rho_{1,2}(t) \\
					\tfrac{1}{\sqrt{2}}\rho_{0,1}^*(t)	&& \tfrac{1}{2}\rho_{1,1}(t)				&& \tfrac{1}{2}\rho_{1,1}(t)			&& \tfrac{1}{2}\rho_{1,2}(t)		&& \tfrac{1}{2}\rho_{1,2}(t) \\
					\tfrac{1}{\sqrt{2}}\rho_{0,2}^*(t)	&& \tfrac{1}{2}\rho_{1,2}^*(t)			&& \tfrac{1}{2}\rho_{1,2}^*(t)		&& \tfrac{1}{2}\rho_{2,2}(t)		&& \tfrac{1}{2}\rho_{2,2}(t) \\
					\tfrac{1}{\sqrt{2}}\rho_{0,2}^*(t)	&& \tfrac{1}{2}\rho_{1,2}^*(t)			&& \tfrac{1}{2}\rho_{1,2}^*(t)		&& \tfrac{1}{2}\rho_{2,2}(t)		&& \tfrac{1}{2}\rho_{2,2}(t) \\
				\end{pmatrix}.
\end{align}
Since $\rho_{1,1}(t)/2$ and $\rho_{2,2}(t)/2$ are just the populations in $|\pm 1,\mp 1\rangle$ and $|\pm 2,\mp 2\rangle$, respectively, we can clearly see that we are guaranteed entanglement between $|1,-1\rangle$ and $|-1,1\rangle$ (and similarly between $|2,-2\rangle$ and $|-2,2\rangle$) provided these populations are non-negligible. Indeed, via postselection we are guaranteed a maximally-entangled state.

Let us consider a concrete example, drawn upon the simulation data reported in the main text. For a relatively low magnetic bias field of $B = 0.2$~G and a temperature of $\SI{8.8}{\micro\kelvin}$, after $\sim 100\,$ms of evolution our simulation predicts a reduced density matrix of
\begin{align}
	\hat{\rho}_\textrm{S}(t)	&= \begin{pmatrix}
					0.5319			&& 0.0002 - 0.0286i		&& 0.0002 - 0.0286i		&& 0.0115 - 0.0014i		&& 0.0115 - 0.0014i \\
					0.0002 + 0.0286i	&& 0.1477   			&& 0.1477  			&& -0.0023 + 0.0031i	&& -0.0023 + 0.0031i \\
   					0.0002 + 0.0286i   	&& 0.1477 			&& 0.1477  			&& -0.0023 + 0.0031i	&& -0.0023 + 0.0031i \\
   					0.0115 + 0.0014i  	&& -0.0023 - 0.0031i  	&& -0.0023 - 0.0031i 	&& 0.0846				&& 0.0846 \\
   					0.0115 + 0.0014i  	&& -0.0023 - 0.0031i  	&& -0.0023 - 0.0031i   	&& 0.0846				&& 0.0846
				\end{pmatrix}.
\end{align}
In accordance with our intuition that spin-changing collisions preserve magnetization, the off-diagional elements corresponding to entanglement between $|0,0\rangle$ and $|-1,1 \rangle$, for instance, are much smaller than those corresponding to entanglement between $|1,-1\rangle$ and $|-1,1 \rangle$. Indeed, if we postselect on the atoms being in the $m_F = \pm 1$ state, then 30\% of the time we generate the maximally-entangled state $\tfrac{1}{\sqrt{2}}(|1,-1\rangle + |-1,1\rangle)$. We emphasize that this entanglement occurs for a thermal ensemble -- that is, it is preserved under incoherent averaging over the motional degrees of freedom.

Furthermore, this entanglement is metrologically useful; as proven in Ref.~~\cite{Pezze:2009}, when a maximally-entangled state such as $\tfrac{1}{\sqrt{2}}(|1,-1\rangle + |-1,1\rangle)$ forms the input of a Ramsey or Mach-Zehnder interferometer it enables Heisenberg-limited sensitivities $\propto 1 / N$. In contrast, the mixture $\tfrac{1}{2}\left(|1,-1\rangle \langle 1,-1| + |-1,1\rangle \langle -1,1|\right)$ allows metrology at sensitivities no better than the shot-noise limit $\propto 1 / \sqrt{N}$. Both results follow from a computation of the quantum Fisher information, which is a necessary and sufficient witness of metrologically-useful multiparticle entanglement.

Experimentally, our observation of near-perfect pair correlations between magnetic sublevels is consistent with entanglement, but by itself is not a sufficient condition for entanglement. However, the observed relaxation of the spin populations to $N_{m_F = 0} = N_{m_F = \pm 1} = N_{m_F = \pm 2} = 1/3$ is intriguing. Since atoms in different internal states can be considered distinguishable, complete incoherent thermalization between $|0,0\rangle$, $|-1,1\rangle$, $|1,-1\rangle$, $|-2, 2\rangle$, and $|2, -2\rangle$ should result in $N_{m_F = 0} = 1/5$ and $N_{m_F = \pm 1} = N_{m_F = \pm 2} = 2/5$. In contrast, if the symmetry of the atom-pair system only allows coupling between $|0,0\rangle$, $\hat{\mathcal{S}}|-1,1\rangle = \tfrac{1}{\sqrt{2}}(|1,-1\rangle + |-1,1\rangle)$, and $\hat{\mathcal{S}}|-2,2\rangle =\tfrac{1}{\sqrt{2}}(|2,-2\rangle + |-2,2\rangle)$, then relaxation within this subspace should yield equal populations of $1/3$ -- consistent with our experimental observations.

\subsection{Supplementary Note 3: Pair correlation between $|+1\rangle$ and $|-1\rangle$ as a function of collision time}
We quantify correlations between $|+1\rangle$ and $|-1\rangle$ by computing the relative number squeezing of the population imbalance, which is $\zeta^2=P_1/[N(P_0+P_1)]$ for our measurement (see Methods), where $P_n$ is the probability of $n$ atoms remaining in the optical tweezer after the ejection. In Supplementary Figure~\ref{fig:S3}, the number squeezing is deduced from the data in Fig.~2c for different collision times (without correcting for detection inefficiency) in units of dB below quantum shot noise (QSN). The relative number fluctuations of $11.9\pm0.3$~dB below QSN stated in the main text is obtained by averaging over the measurement results at the collision times 150, 250, 350, and 500 ms.

%%%%%%%%%%%   Figure S2   %%%%%%%%%%%%
\begin{figure}
	\begin{center}
     \includegraphics[width=0.6\linewidth]{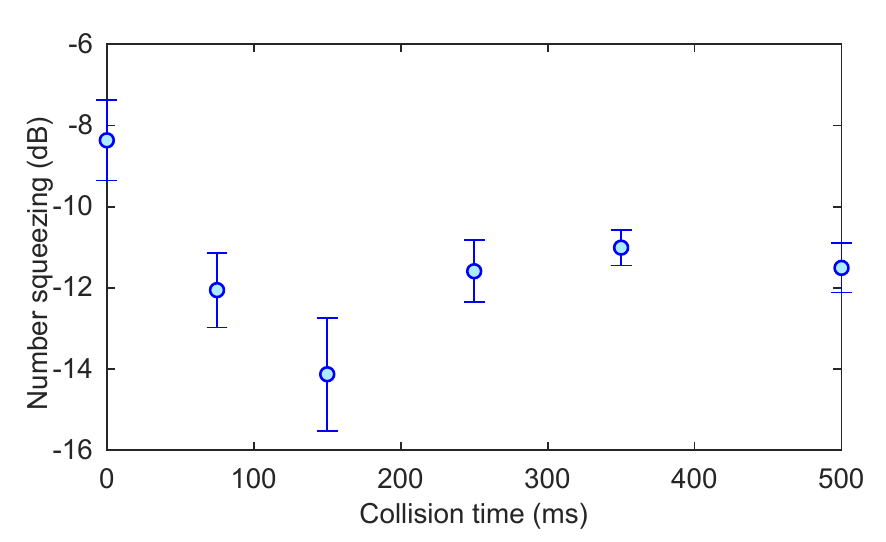}
	\end{center}
   	\caption{
	{\bf Number squeezing.}
	 The relative number squeezing $\zeta^2$ in units of dB below quantum shot noise (QSN) for different collision times, deduced from the data shown in Fig. 2c of the main text. The error bars denote the standard error of the mean $\zeta^2$.}
	\label{fig:S3}
\end{figure}
%%%%%%%%%%%%%%%%%%%%%%%%%%%%%%

\subsection{Supplementary Note 4: Coupling coefficients for spin-exchange collision Hamiltonian}
Consider the spin-changing interaction Hamiltonian
\begin{equation}
	\hat{H}_\textrm{s} = V\left(\hat{\textbf{r}}\right)\sum_{m_1, m_2, m_3, m_4}g_{m_1,m_2}^{m_3,m_4} \left|m_3, m_4 \right\rangle \left\langle m_1,m_2\right|, \label{spin_exchange_Ham}
\end{equation}
where
\begin{equation}
g_{m_1,m_2}^{m_3,m_4}=\sum_{F=0}^{2f}\sum_{M=-F}^{F}g_{F}\left\langle m_3,m_4 |F, M \right\rangle\left\langle F,M|m_1,m_2 \right\rangle,
\label{g_coeffs}
\end{equation}
and $g_F = 4 \pi \hbar^2 a_F / m$ with $a_F$ the $s$-wave scattering length for two atoms colliding in a channel with total spin $F$. Since both atoms are initially prepared in the $m = 0$ Zeeman state, and binary collisions conserve the spin projection along the quantization axis, the summation in Eq.~(\ref{spin_exchange_Ham}) is highly constrained by $m_1 + m_2 = m_3 + m_4 = 0$. Here $|F,M\rangle$ are the eigenstates of the combined Hilbert space of two coupled spins, where $F$ is the total angular momentum quantum number (for two spin-2 atoms, $F = 0, 2, 4$), and $M$ the quantum number associated with the projection onto the quantization axis of this combined space. Consequently, $\langle m_3, m_4 | F, M\rangle$ are Clebsch-Gordon coefficients. These can be taken to be real (so $\langle m_3, m_4 | F, M\rangle = \langle F, M | m_3, m_4 \rangle$), implying that
\begin{equation}
	g_{m_1,m_2}^{m_3,m_4} = g_{m_3,m_4}^{m_1,m_2}.
\end{equation}
Furthermore,
\begin{equation}
	\langle m_1, m_2|F,M\rangle = (-1)^{2f-F} \langle m_2, m_1|F,M\rangle.
\end{equation}
Since $f$ and $F$ are always even, $(-1)^{2f-F} = 1$ always, implying the symmetry
\begin{equation}
	g_{m_1,m_2}^{m_3,m_4} = g_{m_2,m_1}^{m_3,m_4} = g_{m_1,m_2}^{m_4,m_3} = g_{m_2,m_1}^{m_4,m_3}.
\end{equation}
Both permutation symmetries allow us to greatly simplify our expression for $\hat{H}_\textrm{s}$. For example,
\begin{align}
	g_{0,0}^{1,-1} |0, 0\rangle \langle 1, -1| + g_{0,0}^{-1,1} |0, 0\rangle \langle -1, 1|	&= g_{0,0}^{1,-1} |0, 0\rangle \left(  \langle 1, -1| +  \langle -1, 1| \right) \equiv \sqrt{2} g_{0,0}^{1,-1} |0,0\rangle \langle 1,-1| \hat{S},
\end{align}
and
\begin{align}
	&g_{1,-1}^{2,-2} |1, -1\rangle \langle 2, -2| +  g_{-1,1}^{-2,2} |-1, 1\rangle \langle -2, 2| +  g_{1,-1}^{-2,2} |1, -1\rangle \langle -2, 2| +  g_{-1,1}^{2,-2} |-1, 1\rangle \langle 2, -2| \notag \\
	&=  g_{1,-1}^{2,-2} \left( |1, -1\rangle + |-1, 1\rangle \right)\left( \langle2, -2| + \langle-2, 2| \right) \notag \\
	&\equiv 2  g_{1,-1}^{2,-2} \hat{S} |1,-1 \rangle \langle 2,2| \hat{S},
\end{align}
where $\left|0,0\right\rangle$, $\hat{S}\left|1,-1\right\rangle=\frac{1}{\sqrt{2}}\left({\left|1, -1\right\rangle}+{\left|-1, 1\right\rangle}\right)$, and $\hat{S}\left|2,-2\right\rangle=\frac{1}{\sqrt{2}}\left({\left|2, -2\right\rangle}+{\left|-2, 2\right\rangle}\right)$ are the only two-particle spin states accessible by our experiment, due to our choice of $m = 0$ initial condition.

These symmetries and simplifications allow us to write Hamiltonian~(\ref{spin_exchange_Ham}) as
\begin{align}
	\hat{H}_\textrm{s} &= V(\hat{\textbf{r}}) \Big\{  g_{0,0}^{0,0} |0,0\rangle \langle 0,0| + \sqrt{2}  g_{0,0}^{1,-1} \left( |0,0\rangle\langle1,-1|\hat{S} + \hat{S}|1,-1\rangle\langle0,0|\hat{S}\right) + \sqrt{2}  g_{0,0}^{2,-2} \left( |0,0\rangle\langle2,-2|\hat{S} + \hat{S}|2,-2\rangle\langle0,0|\right) \notag \\
			& \qquad\qquad + 2  g_{1,-1}^{1,-1} \hat{S}|1,-1\rangle\langle1,-1|\hat{S} + 2  g_{2,-2}^{2,-2}\hat{S}|2,-2\rangle\langle2,-2|\hat{S} + 2 g_{1,-1}^{2,-2} \left( \hat{S}|1,-1\rangle\langle2,-2|\hat{S} + \hat{S}|2,-2\rangle\langle1,-1|\hat{S}\right) \Big\} \notag \\
			&= V(\hat{\textbf{r}}) \times
			\begin{pmatrix}
				g_{0,0}^{0,0} & \sqrt{2} g_{0,0}^{1,-1} & \sqrt{2}  g_{0,0}^{2,-2} \\
				\sqrt{2}  g_{0,0}^{1,-1} & 2  g_{1,-1}^{1,-1} & 2  g_{1,-1}^{2,-2} \\
				\sqrt{2}  g_{0,0}^{2,-2} & 2  g_{1,-1}^{2,-2} & 2  g_{2,-2}^{2,-2}
			\end{pmatrix} \label{eq_Hs}.
\end{align}
Clearly the spin-changing interaction Hamiltonian is determined by six unique coupling coefficients. It is straightforward to compute these coupling coefficients; since $\langle F, M | m_1; m_2\rangle = 0$ for $m_1 + m_2 \neq M$, we only need to keep terms in Eq.~(\ref{g_coeffs}) where $M = 0$:
\begin{align}
	g_{m_1,m_2}^{m_3,m_4} 	&= g_0 \langle m_3, m_4 | 0, 0\rangle \langle 0, 0 | m_1, m_2\rangle + g_2 \langle m_3, m_4 | 2, 0\rangle \langle 2, 0 | m_1, m_2\rangle + g_4 \langle m_3, m_4 | 4, 0\rangle \langle 4, 0 | m_1, m_2\rangle.
\end{align}
Explicitly evaluating the Clebsch-Gordon coefficients gives Eqs.~(7) in the Methods.

\subsection{Supplementary Note 5: Thermal distribution for two-particle states in a 3D harmonic potential}
Initially, our two atoms are always in the $|0,0\rangle$ state, so $\psi_1(\textbf{r},0) = \psi_2(\textbf{r},0) = 0$ (recall that $\langle \textbf{r} |\psi(t)\rangle = \sum_{m=0,1,2} \psi_m(\textbf{r},t) \hat{S}|m,-m\rangle$). In any given experiment, the two atoms are prepared in a specific eigenstate $\psi_0(\textbf{r},0) = \varphi_{n_x}(x)\varphi_{n_y}(y)\varphi_{n_z}(z)$, but \emph{only} for values of $n_x, n_y, n_z$ where $(-1)^{n_x+n_y+n_z} = 1$ (since $\psi_0(\textbf{r})$ must be symmetric under particle exchange). Here $\varphi_{n_i}(x_i)$ are the eigenstates of the 1D harmonic oscillator of mass $\mu$ and frequency $\omega_i$.

Assuming these constraints on the allowable eigenstates, then within the canonical ensemble (i.e. Boltzmann statistics), the probability that $\psi_0(\textbf{r}, 0)$ will be prepared in the eigenstate with quantum numbers $(n_x, n_y, n_z)$ is
\begin{align}
	\mathcal{P}(n_x,n_y,n_z)	&=  \frac{1}{\mathcal{Z}}\exp\left\{ -\beta \left[ \hbar \omega_x (n_x+\tfrac{1}{2})+\hbar \omega_y (n_y+\tfrac{1}{2})+\hbar \omega_z (n_z+\tfrac{1}{2})\right]\right\},
\end{align}
where $\beta = 1 / k_\textrm{B} T$ and $\mathcal{Z}$ is the partition function, given by the sum over Boltzmann factors for each \emph{allowable} state. This can be written
\begin{equation}
	\mathcal{Z} = \mathcal{Z}_x^\text{even} \mathcal{Z}_y^\text{even} \mathcal{Z}_z^\text{even} + \mathcal{Z}_x^\text{odd} \mathcal{Z}_y^\text{odd} \mathcal{Z}_z^\text{even} + \mathcal{Z}_x^\text{odd} \mathcal{Z}_y^\text{even} \mathcal{Z}_z^\text{odd} + \mathcal{Z}_x^\text{even} \mathcal{Z}_y^\text{odd} \mathcal{Z}_z^\text{odd},
\end{equation}
where
\begin{align}
	\mathcal{Z}_i^\text{even} 	&= \sum_{m_i=0}^\infty e^{- \beta \hbar \omega_i (2 m_i + 1/2)}, \\
	\mathcal{Z}_i^\text{odd} 	&= \sum_{m_i=0}^\infty e^{- \beta \hbar \omega_i [ (2m_i+1) + 1/2]},
\end{align}
are the partition functions corresponding to the even and odd states of a 1D harmomic oscillator of frequency $\omega_i$, respectively. Analytic expressions for these partition functions exist. For the even case, first note that
\begin{equation}
	\mathcal{Z}_i^\text{even} = q_i^{1/4}\left( 1 + q_i + q_i^2 + \cdots \right),
\end{equation}
where we have defined $q_i \equiv \exp(- 2 \beta \hbar \omega_i)$. Clearly $q_i \mathcal{Z}_i^\text{even} = q_i^{1/4}\left( q_i + q_i^2 + q_i^3 + \cdots \right)$, so $\mathcal{Z}_i^\text{even} - q_i Z_i^\text{even} = q_i^{1/4}$, and therefore
\begin{equation}
	\mathcal{Z}_i^\text{even} = \frac{q_i^{1/4}}{1-q_i} = \frac{e^{-\beta \hbar \omega_i / 2}}{1-e^{-2\beta \hbar \omega_i}}.
\end{equation}
Similarly, we can show that
\begin{equation}
	\mathcal{Z}_i^\text{odd} = q_i^{3/4}\left( 1 + q_i + q_i^2 + \cdots \right) = \frac{q_i^{3/4}}{1-q_i} =  \frac{e^{-3\beta \hbar \omega_i / 2}}{1-e^{-2\beta \hbar \omega_i}}.
\end{equation}

\subsection{Supplementary Note 6: Spin-changing collisions for $\delta$-function potential}
Here we derive the relative motion spin-changing evolution equations assuming a $\delta$-function scattering interaction: $V(\textbf{r}) = \delta(\textbf{r})$. We expand $\psi_i(\textbf{r},t)$ in a basis of eigenstates of $H_\text{rel}(\textbf{r}) = -\frac{\hbar^2}{2\mu}\nabla^2_\textbf{r} + \frac{1}{2}\sum_{i=x,y,z}\mu\omega_i^2 r_i^2$:
\begin{equation}
	\psi_i(\textbf{r},t) = \sum_{\textbf{n} \in \mathcal{C}} c_\textbf{n}^i(t) \phi_\textbf{n}(\textbf{r}), \label{psi_exp}
\end{equation}
where $\textbf{n} = (n_x,n_y,n_z)$, $\phi_\textbf{n}(\textbf{r}) = \varphi_{n_x}(x)\varphi_{n_y}(y)\varphi_{n_z}(z)$, $\mathcal{C} = \{ \textbf{n}: \epsilon_\textbf{n} \leq E_\text{cut} \}$, and $\epsilon_\textbf{n} = \hbar \omega_x (n_x + 1/2) + \hbar \omega_y (n_y + 1/2) + \hbar \omega_z (n_x + 1/2)$. That is, we only consider a finite number of modes below some energy cutoff $E_\text{cut}$. The 1D Hermite-Gauss modes $\varphi_{n_i}(x_i)$ satisfy
\begin{equation}
	\left[-\frac{\hbar^2}{2\mu}\frac{\partial^2}{\partial x_i} + \frac{1}{2} \mu \omega_i^2 x_i^2\right] \varphi_{n_i}(x_i) = \hbar \omega_i (n_i + \tfrac{1}{2}) \varphi_{n_i}(x_i),
\end{equation}
and are explicitly given by
\begin{equation}
	\varphi_{n_i}(x_i) = (\sigma_i^2 \pi)^{-1/4}\frac{1}{\sqrt{2^{n_i}n_i!}}H_{n_i}(x_i / \sigma_i) e^{-(x_i/\sigma_i)^2/2},
\end{equation}
where $\sigma_i = \sqrt{\hbar / (\mu \omega_i)}$, and $H_n(x)$ are Hermite polynomials.

Substituting Eq.~(\ref{psi_exp}) into Eqs~(9) from the Methods, multiplying both sides by $\phi_\textbf{m}^*(\textbf{r})$, integrating over space, and then exploiting the orthonormality of the eigenstates, we obtain
\begin{subequations}
\begin{align}
	i \hbar \dot{c}_\textbf{n}^0	&= \epsilon_\textbf{n} c_\textbf{n}^0 + \phi_\textbf{n}(0) \sum_\textbf{m} \phi_\textbf{m}(0) \left[ g_{0,0}^{0,0} c_\textbf{m}^0 + \sqrt{2} g_{0,0}^{1,-1} c_\textbf{m}^1 + \sqrt{2}  g_{0,0}^{2,-2} c_\textbf{m}^2 \right], \\
	i \hbar \dot{c}_\textbf{n}^1	&= \left(\epsilon_\textbf{n} + \hbar q_1 B^2 \right) c_\textbf{n}^1 + \phi_\textbf{n}(0) \sum_\textbf{m} \phi_\textbf{m}(0) \left[\sqrt{2} g_{0,0}^{1,-1} c_\textbf{m}^0 + 2 g_{1,-1}^{1,-1} c_\textbf{m}^1 + 2  g_{1,-1}^{2,-2} c_\textbf{m}^2\right], \\
	i \hbar \dot{c}_\textbf{n}^2	&= \left(\epsilon_\textbf{n} + \hbar q_2 B^2 \right) c_\textbf{n}^2 + \phi_\textbf{n}(0) \sum_\textbf{m} \phi_\textbf{m}(0) \left[\sqrt{2} g_{0,0}^{2,-2} c_\textbf{m}^0 + 2 g_{1,-1}^{2,-2} c_\textbf{m}^1 + 2  g_{2,-2}^{2,-2} c_\textbf{m}^2\right],
\end{align}
\label{eqs_HG}
\end{subequations}
where $\phi_\textbf{n}(0) = \varphi_{n_x}(0)\varphi_{n_y}(0)\varphi_{n_z}(0)$ is given by the simple expression
\begin{equation}
	\varphi_{n_i}^i(0) = \begin{cases}
						\sigma_i^{-1/2} \frac{(-2)^{n_i/2} \sqrt{n_i!}}{\pi^{1/4} (n_i/2)!}, & n_i \text{ even}, \\
						0, & n_i \text{ odd}.
						\end{cases} \label{phi0coeffs}
\end{equation}
Equation~(\ref{phi0coeffs}) implies that $\phi_\textbf{n}(0)$ is only nonzero if $n_x$, $n_y$, and $n_z$ are all even. Therefore, if the two-particle wavefunction for the $|0,0\rangle$ spin state is initially prepared in eigenstate $\phi_\textbf{n}(\textbf{r})$, then coupling to (symmetrized) spin states $\hat{S}|1,-1\rangle$, and $\hat{S}|2,-2\rangle$ only occurs if $n_x$, $n_y$, and $n_z$ are all even. However, even-parity states where, for example, $n_x$ is even and $n_y$ and $n_z$ are odd do not undergo spin-changing collisional dynamics according to this model. For our experiment, this represents a significant fraction of the thermal ensemble: for experimental parameters $T = \SI{44}{\micro\kelvin}$, $\omega_x = 2 \pi \times 8.9$~kHz, $\omega_y = 2 \pi \times 55.5$~kHz, and $\omega_z = 1.01 \omega_y$, we have
\begin{align}
	\mathcal{F} 	&= \text{Fraction of states that do not change spin states under $\delta$-function potential} \notag \\
				&= \frac{\mathcal{Z}_x^\text{odd} \mathcal{Z}_y^\text{odd} \mathcal{Z}_z^\text{even} + \mathcal{Z}_x^\text{odd} \mathcal{Z}_y^\text{even} \mathcal{Z}_z^\text{odd} + \mathcal{Z}_x^\text{even} \mathcal{Z}_y^\text{odd} \mathcal{Z}_z^\text{odd}}{\mathcal{Z}} \notag \\
				&\approx 0.733.
\end{align}
This model is therefore at odds with our experimental observations, which showed spin-changing collisional dynamics leading to a transfer of much more than $70$\% of the $|0,0\rangle$ population to the two-particle spin states $|\pm1,\mp1\rangle$ and $|\pm2,\mp2\rangle$.

We briefly remark that this conclusion remains true when the regularized $\delta$-function potential is used: $V(\textbf{r}) = \delta_\text{reg}(\textbf{r}) \equiv \delta(\textbf{r}) \partial_r r = \delta(\textbf{r}) \left( 1 + x \partial_x + y \partial_y + z \partial_z\right)$, where $\partial_{x_i} \equiv \partial / \partial x_i$.

\subsection{Supplementary Note 7: Width of Gaussian scattering potential}
Our numerical simulations use a normalized Gaussian scattering pseudopotential $V(\textbf{r}) = \exp(-r^2 / 2 w^2) / (2\pi w^2)^{3/2}$ with width $w^2 = (a_0^4 + a_2^4 + a_4^4)/(a_0^2 + a_2^2 + a_4^2)$, where the $a_F$ are the $s$-wave scattering lengths for each total spin-$F$ state. Here we show that this choice of width $w$ gives the same overall low-energy scattering cross section (in free space) as the $\delta$-function potential.

Within the Born approximation, the cross section for the spin-$F$ channel is given by ~\cite{Dalibard:1998}
\begin{align}
	\frac{d \sigma_F}{d \Omega} 	&= \frac{\mu^2}{\hbar^4 q^2} \left| \int d \textbf{r} e^{-i \textbf{q} \cdot \textbf{r}} V_F(\textbf{r}) + \int d \textbf{r} e^{i \textbf{q} \cdot \textbf{r}} V_F(\textbf{r}) \right|^2 \notag \\
							&= \frac{\mu^2}{\hbar^4 q^2} \left| \frac{4 \pi}{q}\int_0^\infty dr \, r \sin(qr) V_F(r) + \frac{4 \pi}{(-q)}\int_0^\infty dr \, r \sin(-qr) V_F(r)\right|^2 \notag \\
							&= \frac{16 \mu^2}{\hbar^4 q^2} \left| \int_0^\infty dr \,r \sin(qr) V_F(r) \right|^2.
\end{align}
Here $V_F(\textbf{r}) = g_F V(\textbf{r})$, $\textbf{q} \equiv \textbf{k} - \textbf{k}_0$ and $q = |\textbf{q}|$, where $\textbf{k}$ and $\textbf{k}_0$ represent the momenta of incoming and outgoing plane waves (before and after scattering, respectively). The second term within the absolute value arises due to our requirement that for bosonic particles, the wavefunction needs to be symmetrized; this term is the same as the first but with $\textbf{q} \to - \textbf{q}$ (i.e. we are enforcing exchange symmetry), and for our radially-symmetric potential only results in an additional factor of 4 out the front. Since
\begin{equation}
	\int_0^\infty dr \,r \sin(qr) V_F(r) = \frac{g_F}{4 \pi}q e^{- w^2 q^2 / 2},
\end{equation}
$g_F = 4 \pi \hbar^2 a_F / m$, and $\mu = m/2$, we can write
\begin{equation}
	\frac{d \sigma_F}{d \Omega} = 4 a_F^2 e^{-w^2 q^2}.
\end{equation}

For a radially-symmetric potential, conservation of energy implies that $|\textbf{k}| = |\textbf{k}_0| \equiv k$. We can therefore write $q$ in terms of $k$ and the angle, $\theta$, between $\textbf{k}$ and $\textbf{k}_0$: $q = 2 k \sin(\theta/2)$.

To determine $\sigma_F$, we need to integrate over all \emph{distinct} final scattering states, parametrized by the solid angle $\Omega$. We must therefore only integrate from $0 \leq \theta \leq \pi/2$, $0 \leq \phi \leq 2 \pi$, since the other half-shell $\pi/2 < \theta \leq \pi$ is an identical set of scattered states (follows from the symmetrization requirement; scattering is invariant under exchange $\textbf{q} \to -\textbf{q}$, or $\theta \to \pi - \theta$). Thus,
\begin{align}
	\sigma_F(k)	&= 4 a_F^2 \times \underbrace{2 \pi}_{\phi \textrm{ integral} } \times \int_0^{\pi/2} d\theta \, \sin \theta e^{-4 w^2 k^2 \sin^2(\theta/2) } \notag \\
			&= 8 \pi a_F^2 \left( \frac{1 - e^{-2 w^2 k^2}}{2 k^2 w^2}\right) \notag \\
			&\approx 8 \pi a_F^2 \left( 1 - w^2 k^2\right),
\end{align}
where the final line is approximately true in the limit of low-energy scattering. The \emph{total} cross section is given by the sum over all spin-$F$ channels:
\begin{align}
	\sigma_\text{tot}(k)	&= \sum_F 8 \pi a_F^2 \left( \frac{1 - e^{-2 w^2 k^2}}{2 k^2 w^2}\right) \approx 8 \pi \left[ \left( \sum_Fa_F^2 \right) - \left( \sum_Fa_F^2 \right) w^2 k^2\right].
\end{align}

Now compare this to the cross section for the $\delta$-function pseudopotential ~\cite{Dalibard:1998}:
\begin{equation}
	\sigma_F(k)	= \frac{8 \pi a_F^2}{1 + k^2 a_F^2} \approx 8 \pi a_F^2 \left( 1 - a_F^2 k^2\right),
\end{equation}
and so
\begin{align}
	\sigma_\text{tot}(k)	&\approx \sum_F 8 \pi a_F^2 \left( 1 - a_F^2 k^2\right) = 8 \pi \left[ \left( \sum_F a_F^2\right) - \left( \sum_F a_F^4 \right)k^2\right].
\end{align}
We therefore match the total cross section in the low-energy regime by choosing the width of our Gaussian as:
\begin{equation}
	w^2 = \frac{\sum_F a_F^4}{\sum_F a_F^2} = \frac{a_0^4 + a_2^4 + a_4^4}{a_0^2 + a_2^2 + a_4^2}.
\end{equation}

\subsection{Supplementary Note 8: Coupling matrix for Gaussian pseudopotential}
As can be seen from Hamiltonian Eq.~(10) in the Methods, for the Gaussian pseudopotential, coupling to different spin states is described by the coupling matrix
\begin{equation}
	[\textbf{T}]_{\textbf{n},\textbf{m}} = \frac{1}{(2\pi w^2)^{3/2}} \mathcal{I}_{n_x,m_x}\mathcal{I}_{n_y,m_y}\mathcal{I}_{n_z,m_z},
\end{equation}
where
\begin{align}
	\mathcal{I}_{n_i,m_i} 	&= \int dx_i \, \varphi_{n_i}(x_i) e^{-x_i^2 / 2 w^2} \varphi_{m_i}(x_i) \notag \\
						&= \left( \pi 2^{n_i+m_i} n_i! m_i! \right)^{-1/2} \sigma_i^{-1} \int dx_i H_{n_i}(x_i/\sigma_i) H_{m_i}(x_i/\sigma_i) e^{-\left( 1 + \frac{\sigma_i^2}{2 w^2}\right)\left(\frac{x_i}{\sigma_i}\right)^2} \notag \\
				&= \left( \pi 2^{n_i+m_i} n_i! m_i! \right)^{-1/2} \int d\tilde x_i H_{n_i}(\tilde x_i) H_{m_i}(\tilde x_i) e^{-2 \alpha_{i}^2 \tilde x_i^2},
\end{align}
where $2 \alpha_{i}^2 \equiv 1 + \sigma_i^2 / (2 w^2)$ and $\sigma_i = \sqrt{\hbar / (\mu \omega_i)}$. From result 7.374.5 of Ref.~\cite{Gradshteyn:2014},
\begin{align}
	\int d\tilde x H_{n}(\tilde x) H_{m}(\tilde x) e^{-2 \alpha^2 \tilde x^2}	&= 2^{\frac{m+n-1}{2}}\alpha^{-m-n-1}(1-2\alpha^2)^{\frac{m+2}{2}}\Gamma\left( \tfrac{m+n+1}{2}\right) F\left( -m,-n; \tfrac{1-m-n}{2}; \tfrac{\alpha^2}{2\alpha^2-1}\right),
\end{align}
if $m+n$ is even. If $m+n$ is odd, then this integral is zero. Here $F(a,b;c;d)$ is a Gauss hypergeometric function. Then
\begin{align}
	\mathcal{I}_{n_i,m_i}	&= \begin{cases}
							\frac{\Gamma\left( \tfrac{m_i+n_i+1}{2}\right)}{\sqrt{2 \pi n_i! m_i!}} \alpha_i^{-m_i-n_i-1}(1-2\alpha_i^2)^{\tfrac{m_i+n_i}{2}}F\left( -m_i,-n_i; \tfrac{1-m_i-n_i}{2}; \tfrac{\alpha_i^2}{2\alpha_i^2-1}\right),	& n_i+m_i \textrm{ even} \\
							0,	& n_i+m_i \textrm{ odd}.
							\end{cases}
\end{align}

\end{widetext}

\end{document}